\documentclass[11pt]{article}
\pdfoutput=1

\usepackage{jheppub} 

\usepackage{latexsym,amsmath,amsfonts,amssymb, tensor}
\usepackage{graphicx}
\usepackage{epsf}
\usepackage{makeidx}
\usepackage{multirow}
\usepackage[all]{xy}
\usepackage{hyperref}
\usepackage{verbatim} 
\usepackage{rotating}
\usepackage{xcolor}
\usepackage{multirow}
\usepackage{bm}
\usepackage{cleveref}
\usepackage{slashed}
\usepackage[normalem]{ulem}

\usepackage{dirtytalk}

\usepackage{graphicx}
\usepackage{latexsym,amsmath,amsfonts,amssymb}
\usepackage{mathrsfs}
\usepackage{bbm}
\usepackage{bm}
\usepackage{subfigure}
\usepackage{paralist}
\usepackage{youngtab}
\usepackage{bclogo}
\usepackage{mathrsfs}

\usepackage{esvect} 

\usepackage{mathtools} 

\usepackage{tikz}
\usepackage{tikz-cd}
\usepackage{diagbox}
\usetikzlibrary{positioning}
\usetikzlibrary{calc}
\usetikzlibrary{decorations.pathreplacing,calligraphy}

\usepackage{xstring}
\usetikzlibrary{decorations.pathmorphing} 
\usetikzlibrary{decorations.markings} 
\usetikzlibrary{snakes}
\usetikzlibrary{arrows} 
\usetikzlibrary{shapes} 
\usetikzlibrary{matrix} 
\usetikzlibrary{positioning} 
\usepackage[english]{babel} 
\usepackage[autostyle]{csquotes}

\tikzstyle{brane}=[draw]
\tikzset{D7/.style={circle, draw=black, inner sep=0pt, fill=white, minimum size=3mm}}
\tikzset{hasse/.style={circle, fill,inner sep=2pt}}
\tikzset{flavour/.style={regular polygon,fill=white,regular polygon sides=4,inner sep=2.5pt, draw}}
\tikzset{gauge/.style={circle, draw,inner sep=2.5pt}}
\tikzset{gaugeb/.style={circle, draw,fill=black,inner sep=2.5pt}}
\tikzset{gauger/.style={circle, draw,fill=cyan,inner sep=2.5pt}}
\tikzset{gaugeg/.style={circle, draw,fill=red,inner sep=2.5pt}}
\tikzset{bd/.style={circle, draw=black, inner sep=0pt, fill=black, minimum size=1mm}}
\tikzset{wd/.style={circle, draw=black, inner sep=0pt, fill=white, minimum size=2mm}}
\tikzset{SUd/.style={circle, draw=black, inner sep=0pt, fill=yellow, minimum size=2mm}}
\tikzset{Dynkin/.style={circle, draw=black, inner sep=0pt, fill=white, minimum size=2mm}}
\tikzstyle{ligne}=[draw, thick] 
\tikzset{doublearrow/.style={ draw=black!75, color=black!75, thick, double distance=3pt, }}

\numberwithin{equation}{section}

\newcommand{\nn}{\nonumber}
\newcommand{\mat}[1]{\begin{pmatrix} #1 \end{pmatrix}}

\newcommand{\be}{\begin{equation}} 
\newcommand{\ee}{\end{equation}}
\newcommand{\bea}{\begin{equation} \begin{aligned}} \newcommand{\eea}{\end{aligned} \end{equation}}

\newcommand{\bit}{\begin{itemize}} 
\newcommand{\eit}{\end{itemize}}

\newcommand{\Z}{\mathbb{Z}}
\newcommand{\C}{\mathbb{C}}
\newcommand{\K}{\mathbb{K}}
\newcommand{\R}{\mathbb{R}}

\renewcommand{\t}{\widetilde }
\renewcommand{\d}{\partial }

\newcommand{\half}{{1\over 2}}

\newcommand{\CB}{\mathcal{B}}

\newcommand{\CG}{\mathcal{G}}
\newcommand{\CH}{\mathcal{H}}
\newcommand{\CI}{\mathcal{I}}

\newcommand{\CN}{\mathcal{N}}
\newcommand{\CO}{\mathcal{O}}

\newcommand{\CR}{\mathcal{R}}
\newcommand{\CS}{\mathcal{S}}
\newcommand{\CT}{\mathcal{T}}

\newcommand{\CV}{\mathcal{V}}
\newcommand{\CW}{\mathcal{W}}

\newcommand{\FR}{\mathfrak{R}}

\newcommand{\n}{\mathfrak{n}}

\newcommand{\h}{\widehat}

\DeclareMathOperator{\Tr}{Tr}
\DeclareMathOperator{\tr}{tr}

\DeclareMathOperator{\sign}{sign}

\newcommand{\ov}{\over}

\newcommand{\dilog}{{\text{Li}_2}}

 \newcommand{\T}{\mathfrak{T}} 
 \newcommand{\TW}{\textbf{T}} 
  \newcommand{\SW}{\textbf{S}} 
 
\DeclareMathAlphabet{\pazocal}{OMS}{zplm}{m}{n}

\usepackage{diagbox}
\usepackage{array}
\makeatletter
\newcommand{\thickhline}{%
    \noalign {\ifnum 0=`}\fi \hrule height 1pt
    \futurelet \reserved@a \@xhline
}
\newcolumntype{"}{@{\hskip\tabcolsep\vrule width 1pt\hskip\tabcolsep}}
\makeatother

\makeatletter
\g@addto@macro{\endtabular}{\rowfont{}}
\makeatother
\newcommand{\rowfonttype}{}
\newcommand{\rowfont}[1]{
   \gdef\rowfonttype{#1}#1%
}
\newcolumntype{L}{>{\rowfonttype}l}

\usepackage{multirow}
\usepackage{arydshln}

\begin{document}

\baselineskip=18pt  
\numberwithin{equation}{section}  
\allowdisplaybreaks  


%
%


\thispagestyle{empty}

\vspace*{0.8cm} 
\begin{center}
{{\Huge  
Twisted indices, Bethe ideals \\
\medskip 
and 3d $\mathcal{N}=2$ infrared dualities
}}

 \vspace*{1.5cm}
Cyril Closset,  Osama Khlaif

 \vspace*{0.7cm} 

 {  School of Mathematics, University of Birmingham,\\ 
Watson Building, Edgbaston, Birmingham B15 2TT, United Kingdom}\\

\vspace*{0.8cm}
\end{center}
\vspace*{.5cm}

\noindent
We study  the topologically twisted index of 3d $\CN=2$  supersymmetric gauge theories with unitary gauge groups. We implement a Gr\"obner basis algorithm for computing the $\Sigma_g\times S^1$ index explicitly and exactly in terms of the associated Bethe ideal, which is defined as the algebraic ideal associated with the Bethe equations of the corresponding 3d $A$-model. We then revisit recently discovered infrared dualities for unitary SQCD with gauge group $U(N_c)_{k, k +l N_c}$ with $l\neq 0$, namely the Nii duality that generalises the Giveon-Kutasov duality,  the Amariti-Rota duality that generalises the Aharony duality, and their further generalisations in the case of arbitrary numbers of fundamental and antifundamental chiral multiplets. In particular, we determine all the flavour Chern-Simons contact terms needed to make these dualities work. This allows us to check that the twisted indices of dual theories match exactly. We also  initiate the study of the Witten index of unitary SQCD with $l\neq 0$.

\newpage

\tableofcontents


\section{Introduction}

Many half-BPS observables of 3d $\CN=2$ supersymmetric field theories can be computed exactly. In particular, exact results are known for the supersymmetric partition functions of 3d $\CN=2$ gauge theories on any half-BPS closed 3-manifold -- see {\it e.g.}~\cite{Kim:2009wb, Kapustin:2009kz, Jafferis:2010un, Imamura:2011su, Kapustin:2011jm} and the review~\cite{Willett:2016adv} for the best studied examples. These exact results are obtained through supersymmetric localisation methods, and they are typically given in terms of an ordinary integral over the Cartan subalgebra of the gauge algebra or of some complexification thereof. This leaves one with the still-challenging task of evaluating that integral explicitly.

An alternative approach uses the Seifert fibration that exists on most half-BPS closed 3-manifolds~\cite{Closset:2012ru} -- see~\cite{Closset:2019hyt} for a review. In this case, the computation of many observables is reduced to a problem in an auxilliary 2d $\CN=(2,2)$ field theory subjected to the topological $A$-twist, sometimes called the 3d $A$-model~\cite{Closset:2017zgf,Closset:2018ghr}.  In this work, we are concerned with the simplest such partition function, the so-called {\it twisted index}, which captures the Witten index~\cite{Witten:1981nf} of the 3d $\CN=2$ gauge theory  on a Riemann surface of genus $g$. It is equal to the supersymmetric partition function on $\Sigma_g\times S^1$ \cite{Nekrasov:2014xaa, Benini:2015noa, Benini:2016hjo, Closset:2016arn}:
\be
Z_{\Sigma_g\times S^1}(y)= {\rm Tr}_{\Sigma_g}\big( (-1)^D y^{Q_F})\big)~.
\ee
The twisted index depends on various fugacities, here denoted by $y$, for the flavour symmetry with charges $Q_F$. For $g=1$, the twisted index reduces to the ordinary (flavoured) Witten index, which must be an integer~\cite{Witten:1999ds, Intriligator:2013lca}. More generally, the twisted index is a rational function of the flavour fugacities:
\be
Z_{\Sigma_g\times S^1} \, \in \, \C(y)~.
\ee
We are interested in computing this index as explicitly and efficiently as possible. It is most convenient to view the index as a trace over the Hilbert space of the 3d $A$-model, which amounts to a sum over all the so-called Bethe vacua (the 2d vacua of the $A$-model)   \cite{Nekrasov:2014xaa}:
\be\label{bethe vac sum intro}
Z_{\Sigma_g\times S^1}(y) = \sum_{\h x\in \CS_{\rm BE}} \CH(\h x, y)^{g-1}~.
\ee
Here $x=\h x$ are the solutions to the Bethe equations (the 2d vacuum equation), which reads:%
\footnote{The name `Bethe equations' is the short form of `Bethe ansatz equations', and it arose because of the Bethe/gauge correspondence of Nekrasov and Shatashvili~\protect\cite{Nekrasov:2009uh}.}
\be
\Pi_a(x)\equiv \exp\left( (2\pi i)^2   {\d \CW\ov \d \log x_a} \right)=1~,\qquad \qquad a=1~, \,\cdots~,\, \text{rk}(G) ~,
\ee
schematically. $G$ denotes the gauge group and $\CW$ is the effective twisted superpotential of the 3d $A$-model. The object $\CH(x,y)$ appearing in~\eqref{bethe vac sum intro} is the handle-gluing operator~\cite{Nekrasov:2014xaa}.  The Witten index of the 3d $\CN=2$ supersymmetric theory is then identified with the number of Bethe vacua in the corresponding 3d $A$-model.%
\footnote{To avoid any confusion, let us recall that, despite its ill-conceived name, the 3d $A$-model is a 2d TQFT.}

In this paper, we choose $G$ to be a product of unitary gauge groups, $G= \prod_{I} U(N_I)$. In this case, we can   implement powerful computational algebraic geometry methods to compute the sum \eqref{bethe vac sum intro}, as advocated by~\cite{Jiang:2017phk, LykkeJacobsen:2018nhn}  in a slightly different context. The basic idea is to view the Bethe equations as generating an algebraic ideal, called the Bethe ideal $\CI_{\rm BE}$, in the polynomial ring $\C[x]$, and to expand this ideal in a Gr\"obner basis $\CG(\CI_{\rm BE})$. Then,  we can associate to the $A$-model operator $\CH$   its  companion matrix~\cite{cox2006using}, a $\C(y)$-valued matrix obtained by expanding out the handle-gluing operator along $\CG(\CI_{\rm BE})$. The twisted index \eqref{bethe vac sum intro} can be computed as the trace of the companion matrix of $\CH$ raised to the $(g-1)^\text{th}$ power. These Gr\"obner-basis computations are easily implemented on a computer, which allows us to evaluate \eqref{bethe vac sum intro} completely explicitly in numerous examples.%
\footnote{All the algebraic geometry computations for this paper were performed using {\sc Singular}~\protect\cite{DGPS}  interfaced with {\sc Mathematica}~\protect\cite{Mathematica} using a modified version of the {\it Singular.m} package~\protect\cite{Singularm}.}  Note that Gr\"obner bases were previously used in~\cite{Eckhard:2019jgg} to compute the Witten index of various theories; see also the recent work~\cite{Gu:2022dac} for a related approach to 3d $\CN=2^\ast$ quivers.

Another reason to study unitary gauge theories is that they enjoy infrared (Seiberg-like) dualities~\cite{Intriligator:1996ex, Aharony:1997gp, Giveon:2008zn, Benini:2011mf}. Such dualities have been extensively studied in recent years~\cite{Niarchos:2008jb, Cremonesi:2010ae, Kapustin:2011gh, Kapustin:2011vz, Aharony:2011ci, Park:2013wta, Aharony:2014uya, Benini:2017dud, Amariti:2017gsm, Dimofte:2017tpi, Hwang:2018uyj, Fazzi:2018rkr, Benvenuti:2018bav, Nii:2018bgf, Nii:2019qdx, Giacomelli:2019blm, Nii:2019wjz, Nii:2020ikd, Kubo:2021ecs, Okazaki:2021gkk, Benvenuti:2020wpc, Hwang:2022jjs, Amariti:2020xqm, Amariti:2022iaz, Amariti:2022lbw}, and they can even be related to non-supersymmetric `bosonisation' dualities in 3d --see  {\it e.g.}~\cite{Aharony:2015mjs, Seiberg:2016gmd, Karch:2016sxi, Gur-Ari:2015pca, Kachru:2016rui, Karch:2016aux, Aharony:2016jvv, Benini:2017dus}. In the second half of this paper, we  focus on SQCD$[N_c, k, l, n_f, n_a]$, defined as a 3d $\CN=2$ $U(N_c)_{k, k+l N_c}$ gauge theory with $n_f$ fundamental chiral multiplets and $n_a$ antifundamental chiral multiplets. Importantly, we allow for generic Chern-Simons levels $k$ and $(k+l N_c)N_c$ for the $SU(N)$ and $U(1)$ factors of  $U(N_c)\cong (SU(N_c)\times U(1))/ \Z_{N_c}$. We initiate a more systematic study of unitary SQCD with $l\neq 0$, including by computing its Witten index when $n_f= n_a$. 

The precise form of the dual gauge theory description of unitary SQCD depends non-trivially on the CS levels $k$ and $l$. For $l=0$, we have the famous Aharony duality (for $k=0$, $n_f=n_a$) \cite{ Aharony:1997gp}, the Giveon-Kutasov duality (for $k\neq 0$, $n_f=n_a$) \cite{Giveon:2008zn}, and some `chiral' Seiberg-like dualities (for $n_f\neq n_a$) \cite{Benini:2011mf}. Very recently, these dualities were generalised to the case $l\neq 0$ by Nii~\cite{Nii:2020ikd} (for $k\neq 0$, $n_f=n_a$) and by Amariti and Rota~\cite{Amariti:2021snj} (in the other cases). All these new dualities can be obtained from the standard $l=0$ dualities by a suitable application of the Kapustin-Strassler-Witten ${\rm SL}(2,\Z)$ action on 3d $\CN=2$ field theories~\cite{Kapustin:1999ha, Witten:2003ya}, as first pointed out in~\cite{Amariti:2021snj}. We revisit this derivation and clarify some subtle aspects of it, especially as it pertains to the various Chern-Simons contact terms which must be determined in order to fully specify any 3d duality. Finally, we can use our formalism to compute the twisted index of dual theories. We verify that they match exactly in many examples, as expected. This is a nice check on our derivation of these dualities with $l\neq 0$, including of all their (background and dynamical) Chern-Simons terms.

\medskip 
\noindent
This paper is organised as follows. In section~\ref{sec:twisted indices}, we set up the stage and discuss the class of 3d $\CN=2$ unitary gauge theories that we will study. After explaining our 3d conventions, we show how to compute the twisted index using the companion matrix algorithm. In section~\ref{sec:index sqcd}, we study some aspects of the twisted index of unitary SQCD, including its Witten index. In section~\ref{sec:SQCDl0}, we review the infrared dualities for SQCD with $l = 0$. Finally, we rederive all the $l\neq 0$ dualities in section~\ref{sec:SQCDlnon0}.

\section{Twisted indices of unitary gauge theories from Bethe ideals}\label{sec:GUnitary and AG}\label{sec:twisted indices}
In this section, we discuss any 3d $\CN=2$ supersymmetric gauge theory with a gauge group:
\be
G= \prod_{I=1}^{n_G} U(N_I)~,
\ee
 with chiral multiplets  $\Phi_i$ in  representations $\FR_i$ of $G$. For any such theory, we explain how to compute the twisted index exactly, using the computational algebraic geometry approach of~\cite{Jiang:2017phk}. For each $U(N)$ gauge group, we keep track of the Chern-Simons levels $K$ and $K+L N$, corresponding to the two factors in $U(N)\cong (SU(N)\times U(1))/\Z_N$. For $A=A_\mu dx^\mu$ a $U(N)$ gauge field, we have:
\be\label{klCS term}
i{K\ov 4\pi}\int {\rm tr}\left(A \wedge dA- {2i\ov 3} A^3 \right)+ i{L\ov 4\pi}\int {\rm tr}(A)\wedge d\, {\rm tr}(A),
\ee
with the trace in the fundamental representation, plus the standard supersymmetric completion. 
  Setting $k=K$ and $l=L$, this Chern-Simons theory is generally denoted by \cite{Hsin:2016blu}:
\be\label{UNkl def}
U(N)_{k,\, k+l N}~,
\ee
and we sometimes use the notation $U(N)_k$ for the special case $l=0$. In this normalisation, the overall $U(1)\subset U(N)$ gauge field ${1\ov N}{\rm tr}(A)$ has an abelian CS level:
\be
k_{U(1)}= N(k+l N)~.
\ee
Let us already note that, in the presence of charged fermions, the levels $k$, $l$ appearing in \eqref{UNkl def} are not exactly the bare CS levels $K$, $L$ that appear in \eqref{klCS term} due to certain one-loop shifts, as we will review momentarily. 

\subsection{Twisted index and Bethe vacua}
The $\Sigma_g$ twisted index is the Witten index of the 3d $\CN=2$ supersymmetric theory compactified on $\Sigma_g$ with the topological $A$-twist \cite{Nekrasov:2014xaa, Benini:2015noa, Benini:2016hjo, Closset:2016arn}:
\be\label{TopIndex0}
Z_{\Sigma_g\times S^1}(y)_\n= {\rm Tr}_{\Sigma_g; \n}\left( (-1)^{\rm F} \prod_\alpha y_\alpha^{Q_F^\alpha} \right)~.
\ee
 We turn on background vector multiplets preserving the two $A$-twist supercharges, hence the index depends on fugacities $y_\alpha$ and on background fluxes $\n_\alpha$. We chose a maximal torus of the flavour group $G_F$, whose rank is denoted by $r_F$:
 \be
 \prod_{\alpha=1}^{r_F} U(1)_\alpha \subseteq G_F~,
 \ee
 and $Q_F^\alpha$ denotes the corresponding conserved charges. 
  We choose the periodic boundary condition along the $S^1$ for the fermions.%
  \footnote{See~\protect\cite{Closset:2018ghr} for a discussion of the more general case.} We also choose the so-called `$U(1)_{-\half}$ quantisation' for all chiral multiplets \cite{Closset:2017zgf, Closset:2018ghr}, as we will review momentarily.

After compactifying  the 3d theory on a circle, let us consider the Coulomb branch of the effective 2d $\CN=(2,2)$ Kaluza-Klein (KK) theory, which is spanned by dimensionless gauge parameters
\be
u_{a_I}= a_{ a_I}^{(0)}+i \beta \sigma_{a_I}~,  
\ee
where $a_{a_I}^{(0)}$ corresponds to the holonomy of the 3d abelian gauge field along the circle of radius $\beta$, and $\sigma_{a_I}$ are the 3d vector-multiplet scalars; here we use the index $a_I=1, \cdots, N_I$ for a given $U(N_I)$ factor. The low-energy abelian gauge group is the maximal torus subgroup:
\be
\prod_{I=1}^{n_G} \prod_{a_I=1}^{N_I} U(1)_{a_I} \subseteq G~,
\ee 
of rank $r_G= \sum_{I=1}^{n_G} N_I$.
The twisted index can be computed entirely in terms of the effective twisted superpotential $\CW$ and the effective dilaton potential $\Omega$. These  holomorphic functions of the gauge parameters $u_{a_I}$ and of  flavour parameters $\nu_\alpha$ govern the topological $A$-twist of the 2d $\CN=(2,2)$ KK theory -- this 2d TQFT is also called the `3d $A$-model' \cite{Closset:2017zgf}. Because of 3d gauge invariance, the variables $u$ and $\nu$ are periodic ($u_{a_I}\sim u_{a_I}+1$ and $\nu_\alpha\sim \nu_\alpha +1$), and it is useful to introduce the single-valued variables:
\be
x_{a_I}\equiv e^{2\pi i u_{a_I}}~, \qquad \qquad
y_\alpha \equiv e^{2\pi i \nu_\alpha}~,
\ee
with the index \eqref{TopIndex0} being a meromorphic function of the flavour fugacities  $y_\alpha$. 
 For each $U(N_I)$ gauge group, we also have a residual gauge symmetry $S_{N_I}$, the Weyl group of $U(N_I)$, which acts as permutations of the $N_I$ parameters $x_{a_I}$ (at any fixed $I$).

\medskip
\noindent
{\bf Effective twisted superpotential.} 
The full  low-energy twisted superpotential of the 3d gauge theory on a circle is given by the sum of some `matter' and Chern-Simon contributions:
\be
\CW= \CW_{\rm matter}+ \CW_{{\rm CS},GG}+  \CW_{{\rm CS},GF} +  \CW_{{\rm CS},FF}~,
\ee
with
\bea\label{W in detail}
& \CW_{\rm matter} &=&\;\; {1\ov (2\pi i)^2} \sum_i \sum_{\rho_i\in \FR_i}  \dilog(x^{\rho_i} y^{\rho_{F, i}})~, \\
&  \CW_{{\rm CS},GG} &=&\;\;  \sum_{I=1}^{n_G}\left( {K_I \ov 2} \sum_{a_I=1}^{N_I} (u_{a_I}^2+ u_{a_I})+ {L_I\ov 2} \left(\left(\sum_{a_I=1}^{N_I} u_{a_I}\right)^2+  \sum_{a_I=1}^{N_I} u_{a_I} \right)\right) \\
&&&\;+ \sum_{I> J}  K_{IJ} \left(\sum_{a_I=1}^{N_I} u_{a_I}\right) \left(\sum_{a_J=1}^{N_J} u_{a_J}\right)~,\\
&  \CW_{{\rm CS},GF} &=&\;\;   \sum_{I=1}^{n_G}   \sum_{\alpha=1}^{r_F} K_{\alpha I} \nu_\alpha  \left(\sum_{a_I=1}^{N_I} u_{a_I}\right)~,\\
&  \CW_{{\rm CS},FF} &=& \sum_{\alpha=1}^{r_F}\;\; {K_{\alpha} \ov 2} (\nu_\alpha^2+\nu_\alpha) + \sum_{\alpha>\beta} K_{\alpha\beta} \nu_\alpha \nu_\beta + {1\ov 24} K_g~.
\eea
Here, we have $x^{\rho_i} \equiv e^{2\pi i \rho_i(u)}$, for any weight $\rho_i$ of the representation $\FR_i$ of $G$ under which the chiral multiplet $\Phi_i$  transforms. We also defined $y^{\rho_{F, i}} \equiv e^{2\pi i \rho_{F, i}(\nu)}$, where $\rho_{F,i}(\nu)=\sum_\alpha \rho^\alpha_{F, i }\nu_\alpha$, in terms of the $U(1)_\alpha$ flavour charges $\rho^\alpha_{F, i}= Q_F^\alpha[\Phi_i]$. 
In addition, the `gauge-gauge' Chern-Simons terms $\CW_{{\rm CS},GG}$ are the bare Chern-Simons terms for the gauge group, including mixed CS terms between distinct $U(N_I)$ factors, and similarly for the mixed gauge-flavour CS terms $\CW_{{\rm CS},GF}$ and the pure flavour CS terms $\CW_{{\rm CS},FF}$, where we included the gravitational CS level $K_g$ \cite{Closset:2012vg}.%
\footnote{The gravitational CS term $K_g$ will not play any role in in this work, but we will keep track of it as it affects the phase of partition functions on generic Seifert manifolds~\protect\cite{Closset:2017zgf, Closset:2018ghr}.} 
Note that $K_{IJ}=K_{JI}$ and $K_{\alpha\beta}=K_{\beta\alpha}$. 
 The Fayet–Iliopoulos terms for each $U(N_I)$ factor appear, in this formalism, as part of  $\CW_{{\rm CS},GF}$, as a mixed term between gauge and topological symmetries. (We usually use the notation $\nu_\alpha=\tau$ for a topological symmetry. Note that, due to the non-zero gauge CS levels, only a subset of the naive $n_G$ topological currents will be independent of the other flavour currents, in general.) Let us also note that it is important to distinguish between `bare CS levels' (denoted here by capital letters $K$  or $L$)  and `effective CS levels' in the UV gauge theory (which we denote by $k$ and $l$) -- see \cite{Closset:2012vp, Closset:2018ghr} and section~\ref{subsec:bareCS} below.

\medskip
\noindent
{\bf Effective dilaton.} Similarly, the effective dilaton potential reads:
\be
\Omega= \Omega_{\rm matter}+\Omega_{{\rm CS}}~.
\ee
Here, the first term includes contributions from the chiral multiplets $\Phi_i$ of $R$-charge $r_i$ and from the $W$-bosons, while the second term captures $U(1)_R$-gauge and $U(1)_R$-flavour mixed CS levels, as well as a $U(1)_R$ CS levels \cite{Closset:2012vg}:
\bea\label{Omega in detail}
& \Omega_{\rm matter} &=&\;\; -{1\ov 2\pi i}\sum_{i}  \sum_{\rho_i\in \FR_i}(r_i-1)  \log(1-x^{\rho_i} y^{\rho_{F, i}})
\\ 
&& &\;\; -{1\ov 2\pi i}\sum_{I=1}^{n_G}\sum_{\substack{a_I, b_I\\ a_I\neq b_I} }  \log(1- x_{a_I}x_{I, b_I}^{-1})~, \\
& \Omega_{\rm CS} &=&\;\;  \sum_{I=1}^{n_G} K_{RI} \sum_{a_I=1}^{N_I} u_{a_I} + \sum_{\alpha=1}^{r_F}K_{R\alpha} \nu_\alpha +\half K_{RR}~.
\eea
As in \eqref{W in detail}, we denote the bare CS levels by a capital $K$.

\medskip
\noindent
{\bf Flux operators and handle-gluing operators.} Given $\CW$ and $\Omega$, one finds the gauge and flavour flux operators:
\be
\Pi_{a_I}(x, y) \equiv \exp\left(2\pi i {\d\CW\ov \d u_{a_I}}\right)~, \qquad\quad
\Pi_{\alpha}(x, y) \equiv \exp\left(2\pi i {\d\CW\ov \d \nu_\alpha}\right)~,
\ee
 respectively, 
and the handle-gluing operator:
\be
\CH(x, y)= e^{2\pi i \Omega} \; \times {\rm det}_{a_I, b_J}\left(\d^2 \CW\ov \d u_{a_I} \d u_{b_J}\right)~.
\ee
More explicitly, we  have:
\be\label{Pialpha explicit}
\Pi_{\alpha}(x, y) = \prod_i \prod_{\rho_i\in \FR_i} \left({1\ov 1-x^{\rho_i} y^{\rho_{F, i}}}\right)^{\rho_{F,i}^\alpha} \; 
\prod_{I=1}^{n_G}\left(\prod_{a_I=1}^{N_I} x_{a_I}\right)^{K_{\alpha I}}
\;  \prod_{\alpha=1}^{r_F} (-y_\alpha)^{K_\alpha}\;  \prod_{\beta\neq \alpha} y_\beta^{K_{\alpha\beta}}~,
\ee
for the flavour flux operators, and:
\bea\label{CH explicit}
&\CH(x, y)&=&\; \prod_{i}\prod_{\rho_i\in \FR_i} \left({1\ov 1-x^{\rho_i} y^{\rho_{F, i}}}\right)^{r_i-1}\; \prod_{I=1}^{n_G} \left(\prod_{a_I \neq b_I} {1\ov 1 -x_{a_I} x_{b_I}^{-1}} \, \left(\prod_{a_I=1}^{N_I} x_{a_I}\right)^{K_{RI}}\right) \\
&&& \, \times \; \prod_{\alpha=1}^{r_F} y_\alpha^{K_{R\alpha}} \,   (-1)^{K_{RR}}\;    {\rm det}({\bf H})~,
\eea
for the handle-gluing operator, with the $r_G\times r_G$ Hessian matrix
\be
 {\bf H}_{a_I, b_J}= \sum_i \rho_{i}^{a_I}\rho_i^{b_J}\, {  x^{\rho_i} y^{\rho_{F, i}}\ov 1- x^{\rho_i} y^{\rho_{F, i}}}
 \, +\, \delta_{IJ} \left( \delta_{a_I b_J} K_I + L_I\right) + K_{IJ}~,
\ee
where $K_{IJ}=0$ if $I=J$. 
Importantly,  \eqref{Pialpha explicit} and \eqref{CH explicit} are rational functions of the variables $x_{a_I}$ and $y_\alpha$. 

\medskip
\noindent
{\bf The Bethe equations.} The gauge flux operators for our unitary gauge theories read:
\bea
&\Pi_{a_I}(x, y)&=&\; \prod_i\prod_{\rho_i\in \FR_i} \left({1\ov 1-x^{\rho_i} y^{\rho_{F, i}}}\right)^{\rho_{i}^{a_I}} \;  (-x_{a_I})^{K_I}\; (-1)^{L_I} \, \left(\prod_{b_I=1}^{N_I} x_{b_I}\right)^{L_I}\\
&&&\, \times \prod_{J\neq I} \left(\prod_{b_J=1}^{N_J} x_{b_J}\right)^{K_{IJ}} \;   \prod_{\alpha=1}^{r_F} y_\alpha^{K_{\alpha I}}~. 
\eea
The `Bethe vacua' are defined as the Coulomb-branch vacua of the 2d $\CN=(2,2)$ KK theory, which are determined from the effective twisted superpotential $\CW$. They correspond to solutions to the Bethe equations,
\be\label{BVE 0}
\Pi_{a_I}(x, y)=1~,\quad \forall I, a_I~,
\ee
which are acted on freely by the Weyl group \cite{Hori:2006dk}.%
\footnote{Fixed points of the Weyl group correspond to would-be 2d vacua with a partially restored non-abelian gauge symmetry. This semi-classical analysis receives quantum correction, and it was convincingly argued in~\protect\cite{Hori:2006dk} that such vacua do not contribute -- {\it i.e.} they are lifted by strong-coupling effects.}
 These Bethe equations can be seen as a coupled system of $r_G$ polynomial equations in the $r_G$ variables $x$.  Each complete Weyl orbit of allowable solutions gives a particular vacuum:
\be\label{SBE def}
\CS_{\rm BE} = \Big\{\h x = (\h x_{a_I}) \; \Big| \;\Pi_{a_I}(\h x, y) =1~, \;\; {\rm and} \;\; \h x_{a_I}\neq \h x_{b_I}~, \forall a_I\neq b_I~,\, \;   \text{for each $I$}\Big\}\Big/W_G~.
\ee
Here $W_G= S_{N_1}\times \cdots \times S_{N_m}$ is the Weyl group of $G$.

\medskip
\noindent
{\bf Twisted index as a sum over Bethe vacua.}
The flavoured Witten index \cite{Intriligator:2013lca} of a 3d $\CN=2$ theory is computed by its partition function on the torus with vanishing background fluxes:
\be
{\bf I}_{W} =Z_{T^2\times S^1}(y)_{\n=0}= {\rm Tr}_{T^2}\left( (-1)^{\rm F} \prod_\alpha y_\alpha^{Q_F^\alpha} \right)~.
\ee
On a flat torus, supersymmetry  prevents any dependence on $y_\alpha$, which only acts as an infrared regulator, and the index is an integer which equals the number of Bethe vacua:
\be
{\bf I}_{W} = \left|\CS_{\rm BE} \right|~.
\ee
The topologically twisted index on $\Sigma_g$ is then a natural generalisation of the 3d Witten index. Using the topological invariance along $\Sigma_g$, it can be computed as a trace over the topological $A$-model Hilbert space -- that is, as a sum over Bethe vacua \cite{Nekrasov:2014xaa, Closset:2016arn}:
\be\label{twisted index sum BV}
Z_{\Sigma_g\times S^1}(y)_\n= \sum_{\h x \in \CS_{\rm BE} } \CH(\h x, y)^{g-1}\, \prod_{\alpha=1}^{r_F} \Pi_{\alpha}(\h x, y)^{\n_\alpha}~.
\ee
This is the formula which we would like to evaluate as explicitly as possible in this work.

\medskip
\noindent
{\bf Example: $U(1)_k$ theory with $n_f$ fundamentals.} As a very simple example to illustrate the above formalism, consider a $U(1)_k$ gauge theory with $n_f$ charged chiral multiplets of charge $1$ and $R$-charge $r$ (with the constraint $k+{n_f\ov 2}\in \Z$). Here, the UV Chern-Simons level is equal to $k$ and therefore the bare CS level is
\be
K = k + {n_f\ov 2}~,
\ee
as we review in the next subsection. 
We also have an $SU(n_f)$ flavour symmetry with fugacities $y_i$, $i=1, \cdots, n_f$, with $\prod_{i=1}^{n_f} y_i=1$, and let us say that the $n_f$ flavours transform in the antifundamental of $SU(n_f)$.  The twisted superpotential then reads:
\be
\CW= {1\ov (2\pi i)^2}\sum_{i=1}^{n_f} \dilog(x y_i^{-1}) + {K\ov 2} (u^2+u)+ \tau u~,
\ee
where we have chosen to set all bare CS terms for the non-gauge symmetries to zero except for a mixed CS term $K_{GT}=1$ between the $U(1)$ gauge symmetry and the topological symmetry $U(1)_T$ (this gives the FI term, with $\tau$ the FI parameter). There is a single Bethe equation,
\be\label{P example 1}
\Pi ={(-x)^K q\ov \prod_{i=1}^{n_f} (1-x y_i^{-1})}= 1 \qquad \Leftrightarrow \qquad  P(x)\equiv \prod_{i=1}^{n_f} (y_i-x)- (-1)^K x^K q=0~,
\ee
with the notation $q=e^{2\pi i \tau}$. There are thus ${\rm max}(n_f, K)$ Bethe vacua, corresponding to the roots $\h x$ of this polynomial $P(x)$. The handle-gluing operator reads
\be\label{H example 1}
\CH(x,y)= \prod_{i=1}^{n_f} (y_i-x)^{1-r} \; \left(K+ \sum_{i=1}^{n_f} {x\ov y_i-x}\right)~.
\ee
For instance, let us choose $k=0$, $n_f=2$ and $r=0$, with $(y_1, y_2)=(y, y^{-1})$. The Bethe vacua then correspond to the two solutions
\be
\h x_\pm = \frac{1-q y+y^2\pm \sqrt{\left(1-q y+y^2\right)^2-4 y^2}}{2 y}~,
\ee
and plugging into 
\be\label{ZHexample 1}
Z_{\Sigma_g\times S^1}(y, q)_0 = \CH(\h x_-)^{g-1} +\CH(\h x_+)^{g-1}~,\qquad \CH(x)= 1-x^2,
\ee
 with $\n=0$, one finds
\be\label{Zg exampls}
Z_{S^2\times S^1}=1~, \qquad Z_{T^2\times S^1}=2~,
\qquad 
Z_{\Sigma_2 \times S^1}= 2-y^2-y^{-2} +2 q (y+ y^{-1})-q^2~,
\ee
for $g=0,1,2$. Given its definition as an index, it is clear that the final result for $Z_{\Sigma_g\times S^1}$ must be a rational function of the flavor fugacities, as is indeed the case here.

\subsection{Parity anomaly, CS contact terms and bare CS levels}\label{subsec:bareCS}
When quantising the various Dirac fermions present in our 3d gauge theories,  it is important to specify how we deal with the corresponding parity anomalies \cite{Redlich:1983dv, Niemi:1983rq,AlvarezGaume:1984nf}. Recall that a `parity anomaly' in three space-time dimensions is a mixed anomaly between 3d parity and (background) gauge invariance. (Here `parity' is really time reversal symmetry in Euclidean signature.) We choose to quantise all the fermions that appear in the free UV description in a gauge-preserving manner, hence generally breaking parity. In this, we follow exactly the conventions explained in Appendix~A of \cite{Closset:2018ghr}.

Let us denote the abelianised symmetries $U(1)_{\bf a}$ (where the index ${\bf a}=(a_I, \alpha)$  runs over gauge and flavour indices, considering a maximal torus of $G\times G_F$ for simplicity) under which chiral multiplets $\Phi_i$  have charges $\rho_i^{\bf a}$. For each chiral multiplet, we use the `$U(1)_{-\half}$ quantization', which corresponds to having the CS contact terms $\kappa_{{\bf ab}}= -\half \rho^{\bf a} \rho^{\bf b}$ -- in particular, for a chiral multiplet coupled to a single $U(1)$ with charge $1$, we would generate a CS contact term $\kappa=-\half$, hence the name. We then have the total `matter' one-loop contributions:
\be\label{kappa matter 1}
\kappa_{{\bf ab}}^{\Phi}= -\half \sum_i \sum_{\rho_i\in \FR} \rho_i^{\bf a} \rho_i^{\bf b}~,
\ee
with $\rho_i^{\bf a}=(\rho^{a_I}_i, \rho_{F, i}^\alpha)$, noting that the sum  $\sum_{\rho_i\in \FR}$ only runs over the gauge (non-flavour) weights. A similar discussion holds for non-abelian groups. 

We should also consider the Chern-Simons contact terms involving the $R$-symmetry current and its superpartners \cite{Closset:2012vg,Closset:2012vp}. For the gauginos in vector multiplets, we choose the `symmetric' quantisation for any pair of non-zero roots $\{\boldsymbol{\alpha}, -\boldsymbol{\alpha}\}$, so that the CS contact terms for the gauge symmetry is not shifted. On the other hand, it is most convenient to choose a `$U(1)_{\half}$ quantisation' for the $U(1)_R$ symmetry and for the gravitational CS contact term, which contribute $\delta \kappa_{RR}= \half {\rm dim}(G)$ and $\delta \kappa_g= {\rm dim}(G)$, respectively \cite{Closset:2018ghr}. In summary, we have the UV `matter' contributions \eqref{kappa matter 1} and
\bea\label{kappa matter 2}
&\kappa_{{\bf a} R}^{\Phi}&=&\; -\half \sum_i \sum_{\rho_i\in \FR} \rho_i^{\bf a} (r_i-1)~,\\
&\kappa_{RR}^{\Phi}&=&\;-\half \sum_i {\rm dim}(\FR_i)  (r_i-1)^2 + \half {\rm dim}(G)~,\\
&\kappa_{g}^{\Phi}&=&\;- \sum_i {\rm dim}(\FR_i)    +  {\rm dim}(G)~,\\
\eea
for the gauge-$R$, $RR$ and gravitational contact terms, respectively.

To define the gauge theory in the UV, we need to specify all the `Chern-Simons terms' for   the gauge and  flavour symmetries. This is equivalent to specifying the value of the CS contact terms $\kappa \equiv \kappa^\Phi + K$, namely:
\be\label{kappa split def}
\kappa_{\bf ab} = \kappa^\Phi_{\bf ab} + K_{\bf ab}~,\qquad
  \kappa_{{\bf a}R} = \kappa^\Phi_{{\bf a}R} + K_{{\bf a}R}~, \qquad
 \kappa_{RR} = \kappa^\Phi_{RR} + K_{RR}~, \qquad
  \kappa_g= \kappa^\Phi_g + K_g~,
\ee
where $\kappa^\Phi$ denotes the matter (and gaugino) contributions \eqref{kappa matter 1}-\eqref{kappa matter 2}, and $K$ denotes the `bare CS levels' which appear in the classical Lagrangian. These are the same bare CS levels that appear in \eqref{W in detail} and \eqref{Omega in detail}. Note that, while our choice of quantisation is conventional, the quantities $\kappa$ truly define the UV theory. All the bare CS levels $K$ are integer-quantised, as required by gauge invariance. 

Let us emphasise  that we denote by $\kappa$ the CS contact terms as computed in the free UV theory, before turning on the gauge couplings. As the gauge theory flows to the strongly-coupled infrared, the CS contact terms $\kappa$ for {\it non-gauge} symmetries become non-trivial observables, $\kappa(\mu)$, the parity-odd contributions to two-point functions of conserved currents \cite{Closset:2012vp}. These observables can be computed at the IR fixed point, in principle, by studying  the $S^3$ partition function as a function of the background fields~\cite{Closset:2012vg, Closset:2012ru}.

\medskip
\noindent
{\bf  Chern-Simons levels.}
In keeping with common convention, we shall denote by:
\be
k \equiv \kappa = \kappa^\Phi +K,
\ee
the CS contact terms for the gauge groups in the UV, which we also call the (UV effective) Chern-Simons levels.%
\footnote{In most of the supersymmetric literature, $k$ is simply called `the Chern-Simons level', and we will use this phrase when no confusion is possible. For our purposes, however, it is important to clearly distinguish between the effective CS levels in the UV ($k$) and the bare levels ($K$) that corresponds to terms in the classical  Lagrangian. This is necessary in order to remove any ambiguity (including any sign ambiguity) in the computation of the twisted index, and of all supersymmetric partition function for that matter~\protect\cite{Closset:2019hyt}.} Note that  $k$, unlike $K$, can be half-integers. For each $U(N_I)$ gauge group, we have the levels $k$ and $l$ defined as in \eqref{UNkl def}, with:
\be
k_I = \kappa_I^\Phi + K_I~, \qquad\qquad  l_I= \kappa_{U(1), I}^{\Phi}+ L_I~.
\ee
  We can also have mixed CS levels between different $U(N_I)$ factors, with:
\be
k_{IJ} = \kappa_{IJ}^\Phi + K_{IJ}~, \qquad I\neq J~.
\ee
Decomposing the representations $\FR_i$ (restricted to $U(N_I)$ as appropriate) into $SU(N_I)$ representations $\h \FR_i$ together with $U(1)$ charges $Q_{I, i}$, we have the one-loop contributions:
\bea
& \kappa_I^\Phi= -\half \sum_i T(\h \FR_i)~,\qquad \qquad
 && \kappa_{U(1), I}^{\Phi}= -\half \sum_i  {1\ov N_I^2}  \left({\rm dim}(\FR_i)\,Q_{I,i}^2- N_I\,  T(\h \FR_i) \right)~,\\
&  \kappa_{IJ}^\Phi = -\half \sum_{i} {{\rm dim}(\FR_i)\ov N_I N_J} Q_{I, i} Q_{J, i}~,
\eea
where the quadratic index $T(\h \FR_i)$ for $SU(N_I)$ is normalised so that $T({\rm fund})=1$, and $Q_I({\rm fund})=1$ for the fundamental representation of $U(N_I)$.%
\footnote{Given the weights $\rho_i= (\rho_i^{a_I})$ of the $U(N_I)$ representation $\FR_i$, we have $Q_{I,i}=\sum_{a=1}^{N_I}\rho_i^{a_I}$.}
 Therefore $T({\rm adj})=2N_I$ and $Q_I({\rm adj})=0$ (hence $\delta\kappa_{U(1), I}^{\Phi}=1$) for the adjoint representation of $U(N_I)$. For later purpose, let us also mention that $T({\rm det})=0$ and $Q_I({\rm det})=N_I$ (hence $\delta\kappa_{U(1), I}^{\Phi}=-\half$)  for the 1-dimensional determinant representation.

\subsection{Bethe ideal and companion matrix method}\label{subsec:Grob tech}
While the sum-over-Bethe-vacua formula \eqref{twisted index sum BV} is elegant and simple-looking, to evaluate it explicitly seems to require rather complicated algebraic manipulations, wherein we would first solve the algebraic equations \eqref{BVE 0} in the $r_G$ variables $x$, and then plug the solutions back into the summand appearing in \eqref{twisted index sum BV}. Of course, this is not actually doable except in the very simplest cases, such as for the example \eqref{ZHexample 1}.

In turns out, however, that it is not necessary to explicitly find the solutions $\h x$ in order to compute the twisted index. Instead, one can use powerful computational algebraic geometry methods~\cite{Jiang:2017phk, cox2006using}, as we now review.

\medskip
\noindent
{\bf Companion matrix method.} First, consider any ideal $\CI$ of a polynomial ring in $n$ variables over a field $\K$, $\CI \subset \K[x_1, \cdots, x_n]$. Concretely, it will be generated by a set of $m$ polynomials $P_i(x)$,
\be
\CI= (P)= (P_1, \cdots, P_m)~.
\ee
We also define the associated quotient ring $\CR$ and the algebraic variety $\CV$:
\be
\CR= {\K[x_1, \cdots, x_n]\ov \CI}~, \qquad\qquad  \CV= {\rm Z}(\CI) \cong {\rm Spec}\, \CR~.
\ee
The variety $\CV$ is the set of solutions to the coupled equations:
\be\label{eqs P gen}
P_i(x_1, \cdots, x_n)=0~,\qquad \qquad i=1, \cdots, m~.
\ee
We assume that this variety is {\it zero-dimensional} -- that is, it consists of discrete points $\h x \in \K^n$. This is equivalent to $\CR$ being finite as an abelian group, so let us denote by
\be
d_\CR= |\CR|~,
\ee
the number of discrete solutions. Now, consider two polynomials, $Q_1, Q_2\in \K[x_1, \cdots, x_n]$. We are interested in the quantity
\be\label{ZQ1Q2}
Z(Q_1/Q_2)\equiv \sum_{\h x \in \CV} \, {Q_1(\h x)\ov Q_2(\h x)}~,
\ee
assuming that $Q_2\notin \CI$. 
This can be computed as follows. First, one needs to pick an ordering $\prec$ for the monomials of the polynomial ring $\K[x_1, \cdots, x_n]$. Then, any polynomial $Q$ has a unique leading term, ${\rm LT}(Q)$,  with respect to that ordering. We then choose a Gr\"obner basis for the ideal $\CI$ with respect to the ordering $\prec$, which consist of some $m'$ elements:
\be\label{GB full}
\CG(\CI) = \{g_{1}, \cdots, g_{m'}\}~.
\ee 
A Gr\"obner basis $\CG(\CI)$ is a generating set for $\CI$ such that the leading term of any $P\in \CI$ is proportional to the leading term of some $g_i \in \CG(\CI)$, namely ${{\rm LT}(P)\ov {\rm LT}(g_i)}\in \K$. This then allows one to reduce any polynomial $Q\in \K[x_1, \cdots, x_n]$ along $\CI$, 
\be\label{Q exp def R}
Q(x)= \sum_{i=1}^{m'} c_i \,  g_i(x)\, + R(x)~,\qquad\qquad  c_i \in \K~,
\ee 
where $R(x)$ is called the remainder. Let $[Q]\in \CR$ denote the equivalence class of any polynomial $Q \in \K[x_1, \cdots, x_n]$ in the quotient ring $\CR=\K[x_1, \cdots, x_n]/ \CI$.  Given a Gr\"obner basis, the remainder $R$ in \eqref{Q exp def R} is unique, and thus provides a useful representative of $[Q]=[R]$. Given two polynomials $Q_1$, $Q_2$, we have that $[Q_1]=[Q_2]$ if and only if $R_1=R_2$, and in particular $[Q] =0$ ($Q \in \CI$) if and only if $R=0$.

The determination of Gr\"obner bases can be done on a computer using standard algorithms, implemented most easily using {\sc Singular} \cite{DGPS}. Given a Gr\"obner  basis \eqref{GB full}, we also obtain a canonical $\K$-basis for the quotient ring $\CR$:
\be\label{Kbasis full}
\CB(\CR) = \{\mathfrak{e}_{1}, \cdots, \mathfrak{e}_{d_\CR}\}~.
\ee 
Then,  to any  polynomial $Q\in \K[x_1, \cdots, x_n]$, we can associate a $d_\CR\times d_\CR$ {\it companion matrix} $\mathfrak{M}_Q$ valued in $\K$, which is defined by:
\be
[Q]  [\mathfrak{e}_s] =\sum_{r=1}^{d_\CR} (\mathfrak{M}_Q)_{s r}\,  [\mathfrak{e}_r]~.
\ee
One can show that the companion matrix respects the ring structure, with the product given by matrix multiplication:
\be
\mathfrak{M}_{Q_1+Q_2}=  \mathfrak{M}_{Q_1}+ \mathfrak{M}_{Q_2}~, \qquad\qquad
\mathfrak{M}_{Q_1 Q_2}=  \mathfrak{M}_{Q_1}\, \mathfrak{M}_{Q_2}~.
\ee
We can further generalise this construction to define the companion matrix of rational functions ${Q_1\ov Q_2}\in \K(x_1, \cdots, x_n)$ using the matrix inverse of the denominator:
\be
\mathfrak{M}_{Q_1 /Q_2}=  \mathfrak{M}_{Q_1}\, (\mathfrak{M}_{Q_2})^{-1}~.
\ee
Finally, the key result is that the quantity \eqref{ZQ1Q2} is simply given by the trace of the corresponding companion matrix:
\be
Z(Q_1/Q_2) =\Tr \left(\mathfrak{M}_{Q_1/Q_2}\right)~.
\ee
In fact,  the eigenvalues $\lambda_1, \cdots, \lambda_{d_\CR} \in \K$ of the companion matrix $\mathfrak{M}_Q$ are exactly equal to $Q$ evaluated on the variety $\CV$ \cite{cox2006using}, namely $\lambda_s = Q(\h x_s)$ for some ordering of the solutions $\h x_s$, $s=1, \cdots, d_{\CR}$, to \eqref{eqs P gen}.

\medskip
\noindent
{\bf Application to abelian gauge theories.} The companion matrix method is directly applicable to abelian gauge theories, namely for  $N_I=1$, $\forall I$. The assumption that the variety $\CV$ is zero-dimensional is essentially an assumption that we can lift any non-compact branch of the 3d moduli space of vacua by generic mass deformations -- this is generally possible only if the theory has enough flavour currents. In that case, the Bethe vacua are determined entirely by the conditions:
\be
\Pi_I(x)\equiv {p_{I,1}(x)\ov p_{I,2}(x)}=1  \qquad \Leftrightarrow \qquad  P_I(x)\equiv p_{I,1}(x)- p_{I,2}(x)=0~, \qquad I=1, \cdots, n_G~.
\ee
The polynomials $P_I\in \K[x_1, \cdots, x_{n_G}]$ generate the {\it Bethe ideal}, $\CI_{BE}$. Here,  $\K$ is taken to be
\be\label{field Ky}
\K= \Z(y_1, \cdots, y_{r_F})~,
\ee
 the field of fractions in the flavor fugacities. We can then compute the twisted index \eqref{twisted index sum BV}  in terms of companion matrices, as: 
 \be\label{twisted index sum BV from M}
Z_{\Sigma_g\times S^1}(y)_\n= \Tr\left((\mathfrak{M}_\CH)^{g-1}\, \prod_{\alpha=1}^{r_F} (\mathfrak{M}_{\Pi_\alpha})^{\n_\alpha}\right)~.
\ee
 
\medskip
\noindent
{\bf Simple abelian example:} Let us illustrate the procedure with the example \eqref{P example 1}-\eqref{H example 1} with $K=1$, $n_f=2$ and $r=0$. In that case, the Gr\"obner basis is simply $\{g_1\}=\{x^2 y+x \left(q y-y^2-1\right)+y\}$ and the $\K$-basis is $\{\mathfrak{e}_1,\mathfrak{e}_2\}= \{x,1\}$.
Then, the companion matrix for the handle-gluing operator is
\be
\mathfrak{M}_\CH = \left(
\begin{array}{cc}
  -q^2 -y^2-y^{-2}+2 q (y+y^{-1}) &\quad q- y - y^{-1}\\
-q +y +y^{-1} & 2 \\
\end{array}
\right)~,
\ee
which reproduces \eqref{Zg exampls}. We can similarly compute the companion matrix for the $SU(2)\times U(1)_T$  flavour flux operators:
\be
\mathfrak{M}_{\Pi_{(y)}} = q^{-1} \mat{q y^{-2}+ y^{-1}-y^{-3}\,  &1- y^{-2}\\ -1+ y^{-2} & q+y^{-1} - y}~, \qquad
\mathfrak{M}_{\Pi_{(\tau)}} =\mat{-q + y+y^{-1}\; \;& 1 \\ -1 &0}~.
\ee

\medskip
\noindent
{\bf Bethe ideal for the non-abelian theory.}  Given the gauge group $G=\prod_{I=1}^{n_G} U(N_I)$, the Bethe vacua are given by \eqref{SBE def}. Let us write the Bethe equations in terms of polynomials in the variables $x=(x_{a_I})$ over the field \eqref{field Ky}, as in the abelian case:
\be
\Pi_{a_I}(x)\equiv {p_{a_I,1}(x)\ov p_{a_I,2}(x)}=1~,  \qquad   \qquad P_{a_I}(x)\equiv p_{a_I,1}(x)- p_{a_I,2}(x)=0~.
\ee
Due to gauge invariance, the Bethe equations and all the  flavor flux and handle gluing operators, are symmetric under $S_{N_I}$, the permutation of the variables $x_{a_I}$ for each $I$. Hence we can restrict our attention to Weyl-symmetric polynomials. In particular, we have
\be\label{P in KxW}
P_{a_I} \in \K[x]^{W_G}~.
\ee
In order to discard the spurious solutions to the Bethe equations $P_{a_I}(x)=0$, which correspond to $\h x_{a_I}=\h x_{b_I}$ for $a_I\neq b_I$ at fixed $I$, we use a symmetrisation trick \cite{Jiang:2017phk}. Let us define the polynomials:
\be
\h P_{a_I b_I}(x)\equiv {P_{a_I}- P_{b_I}\ov x_{a_I}- x_{b_I}}~, \qquad a_I > b_I~,
\ee
for  $I=1, \cdots, n_G$.  The Bethe ideal in the $x_{a_I}$ variables is given by the ideal generated by the polynomials $P$ and $\h P$:
\be
\CI_{\rm BE}^{(x)} = (P, \h P)\subset \K[x]^{W_G}~.
\ee
To obtain the Bethe vacua, we should gather the solutions into Weyl orbits, which have size $|W_G|=\prod_{I=1}^{n_G} N_I!$. To avoid this large redundancy, it is convenient to introduce new variables $s_{I, I_a}$ defined as the symmetric polynomials in $x_{a_I}$ (at fixed $I$). More precisely, let us introduce the polynomials
\be
\h S_{a_I}(x, s)\equiv S_{I, a_I}(x) -s_{I, a_I},
\ee
which are linear in $s_{a_I}$. Here, at fixed $I$,  $S_{I, a_I}(x)= S_{a}(x)$ is the $n$-th elementary symmetric polynomial in $N$ variables $x_b$ ($b=1, \cdots, N=N_I$):
\be
S_a(x_1, \cdots, x_N)\, = \sum_{1\leq b_1 < \cdots < b_a \leq N} x_{b_1}\cdots x_{b_a}~.
\ee
Let us also introduce some auxiliary variables $w_I$ and define:
\be
\h W_I\equiv w_I s_{I, N_I} -1~,
\ee
so that imposing $\h W_I=0$ implies $s_{I, N_I}= \prod_{a_I=1}^{N_I} x_{a_I} \neq 0$, thus removing any spurious solutions on which some $x_{a_I}$ variables would vanish (such solutions would be located at infinity on the 3d Coulomb branch). Starting with the extended ideal:
\be
\CI_{\rm BE}^{(x,w,s)}= (P, \h P, \h S, \h W) \subset \K[x, s, w]~,
\ee
we can reduce it to an ideal in $\K[s]$, eliminating $x_{I_a}$ (and $w_I$) from the description by an appropriate choice of monomial ordering:
\be
\CI_{\rm BE}^{(s)}= \CI_{\rm BE}^{(x,w,s)}\big|_\text{reduce}~.
\ee
 We also write any rational operator $\CO(x)$ on the 2d Coulomb branch in terms of the $s_{a_I}$ variables:
\be
\CO(x)\equiv {Q_1(x)\ov Q_2(x)}~, \quad Q_1, Q_2 \in \K[x] \qquad \longrightarrow\qquad 
\CO(s)\equiv {\t Q_1(s)\ov \t Q_2(s)}~,\quad \t Q_1, \t Q_2 \in \K[s]~.
\ee
Let us denote the quotient ring and the variety relative to the Bethe ideal by:
\be
\CR_{\rm BE}^{(s)}= {\K[s]\Big/ \CI_{\rm BE}^{(s)}}~,\qquad \qquad \CV_{\rm BE} \cong {\rm Spec} \, \CR_{\rm BE}^{(s)}~.
\ee
We assume that the `Bethe variety'  $\CV_{\rm BE}$ is zero-dimensional, so that the number of points in $\CV_{\rm BE}\subset \K^{r_G}$  is equal to the number of Bethe vacua:
\be
 \left| \CV_{\rm BE} \right|= d_{\CR_{\rm BE}}= \left| \CS_{\rm BE} \right|~.
\ee
Finally, we choose a Gr\"obner basis $\CG\big(\CI_{\rm BE}^{(s)}\big)$, so that we can define the companion matrix $\mathfrak{M}_\CO$ of any rational operator $\CO(s)$. We can then compute the twisted index exactly as in \eqref{twisted index sum BV from M}.  More generally, the expectation value of any rational operator $\CO$ on $\Sigma_g$ is given by:
\be
\left\langle \CO \right\rangle_{\Sigma_g\times S^1} =\Tr\left((\mathfrak{M}_\CH)^{g-1}\, \mathfrak{M}_\CO \right)~.
\ee 
It is interesting to note that not every 3d $A$-model observable is rational. In particular, the Seifert fibering operators defined in \cite{Closset:2017zgf, Closset:2018ghr} are not rational in $x_{a_I}$ -- instead, they are locally holomorphic functions in the variables $u_{a_I}= {1\ov 2\pi i}\log(x_{a_I})$. It would be very interesting, but likely quite challenging, to extend the methods of this paper to include fibering operators, perhaps using ideas from~\cite{Jiang:2021krx}.

\section{Twisted indices for unitary SQCD}\label{sec:index sqcd}
The formalism of section~\ref{sec:GUnitary and AG} allows us  to compute the twisted index of any unitary gauge theory, in principle.%
\footnote{In practice, we face limitations are due to computing power: to find Gr\"obner basis for large and complicated ideals can be prohibitive on a laptop computer (especially for $\C(y)$-valued polynomials with many distinct $y_\alpha$'s).}
 In the rest of this paper, we will focus on unitary SQCD$[N_c, k, l, n_f, n_a]$, a 3d $\CN=2$ supersymmetric gauge theory with a single unitary gauge group $U(N_c)$ coupled to  $n_f$ fundamental and $n_a$ antifundamental chiral multiplets:
\be\label{SQCD def 0}
U(N_c)_{k, k+l N_c}~,\,  (n_f \,{\tiny\yng(1)}~,\, n_a\,  \overline{ {\tiny\yng(1)}})~,
\ee
 with $k+\half(n_f+n_a)\in \Z$. Interestingly, these theories admit infrared-dual descriptions akin to Seiberg dualities \cite{Seiberg:1994pq}. The case $l=0$ is well understood \cite{Aharony:1997bx, Aharony:1997gp, Giveon:2008zn,Benini:2011mf}, and the general case with $l\neq 0$ has been addressed very recently in the literature \cite{Nii:2020ikd, Amariti:2021snj}. Here we compute the twisted index of these theories, for any values of the parameters. In particular, we compute the Witten index for SQCD with $l\neq 0$ and $n_f=n_a$, which appears to be a new result. We also briefly discuss how Witten indices for different numbers of flavours are related. Other aspects of the vacuum structure of these theories (and the computation of the Witten index for $n_f\neq n_a$ and $l\neq 0$) will be addressed in future work~\cite{toappear2023}.

\subsection{Defining 3d SQCD: Flavour symmetry and Chern-Simons contact terms}\label{subsec:sqcd flavour def}
To fully define the `electric' theory \eqref{SQCD def 0}, we need to specify all the Chern-Simons levels, including the Chern-Simons contact terms for the flavour symmetry, as reviewed in section~\ref{subsec:bareCS}.  The theory has a flavour symmetry:%
\footnote{Perhaps up to a discrete quotient. Here we make no claim about the exact global form of the flavour symmetry group, which could depend in subtle ways on the CS levels -- see~{\it e.g.}~\protect\cite{Cordova:2017kue, Bhardwaj:2022dyt} for related discussions. We thank M.~Bullimore for pointing this out.}
\be
G_F= SU(n_f)\times SU(n_a)\times U(1)_A\times U(1)_T~,
\ee
and a $U(1)_R$ symmetry under which all (anti)fundamental chiral multiplets are  assigned $R$-charge $r$, assuming that $n_a n_f >0$. We will assume that $r\in \Z$ in this paper, so that the theory can be coupled to $\Sigma_g\times S^1$ with the $A$-twist.%
\footnote{We can choose any $r\in \R$ in the UV, and the choice $r\in \Z$ allows us to define the 3d $A$-model on curved space. Whenever the theory flows to a 3d $\CN=2$ SCFT in the IR, there also exists a dynamically determined superconformal $R$-charge, $R_{\rm SCFT}$, which can be computed by $F$-maximisation~\protect\cite{Jafferis:2010un, Closset:2012vg}.} If either $n_f$ or $n_a$ vanishes, we loose the axial symmetry $U(1)_A$ which rotates both fundamental and antifundamental chiral multiplets with the same phase. 
 
 Denoting by $Q_i$, $i=1, \cdots, n_f$, and $\t Q^j$, $j=1, \cdots, n_a$, the fundamental and antifundamental chiral multiplets, we have the charge assignment shown in table~\ref{tab:SQCD charges}. 
\begin{table}[t]
\renewcommand{\arraystretch}{1.1}
\centering
\be\nn
\begin{array}{|c|c|ccccc|}
\hline
    &  U(N_c)& SU(n_f) & SU(n_a)  & U(1)_A &  U(1)_T & U(1)_R  \\
\hline
Q_i        &{\tiny\yng(1)} & \overline{\tiny\yng(1)} & \bm{1}& 1   & 0   &r \\
\tilde{Q}^j   & \overline{\tiny\yng(1)} & \bm{1} &{\tiny\yng(1)}  & 1   & 0   &r \\
\hline
\end{array}
\ee
\caption{Charge assignments for 3d SQCD$[N_c, k, l, n_f, n_a]$.}
\label{tab:SQCD charges}
\end{table}
 No fundamental field is charged under the topological symmetry, $U(1)_T$. The charged objects are the (bare) monopole operators of minimal magnetic flux, $\T^\pm$, which carry topological charge $\pm 1$. The monopole operators carry an electric charge:%
 \footnote{More precisely, they transform into some $Q$-symmetric product of the fundamental representation of $U(N_c)$.}
 \be
 Q_0[\T^\pm]= \pm ( k + l N_c) -  \half (n_f-n_a)~.
 \ee
under the $U(1)\subset U(N_c)$, and they also transform in the representation
\be
\h \FR[\T^\pm] = {\rm Sym}^{\pm k -\half(n_f-n_a)}\left({\tiny\yng(1)}\right)
\ee
of $SU(N_c)$ \cite{Kapustin:2006pk, Benini:2011cma}. The 3d classical Coulomb branch is then lifted by the (effective) CS interactions, in general. We have a 3d quantum Coulomb branch only when $\T^+$ and/or $\T^-$ are gauge-invariant, in which case their VEVs  span the Coulomb branch.

To fully specify the gauge theory, we not only need to specify the $U(N_c)$ CS levels $k$ and $l$, but also any potential mixed CS level between $U(1)\subset U(N_c)$ and the abelian flavour symmetries. Let us first note that we have the one-loop contributions:
\be
\kappa^\Phi_{GG} = -\half (n_f+n_a)~, \qquad \kappa^\Phi_{GA} = -\half (n_f-n_a)~, \qquad  \kappa^\Phi_{GR} =  -\half (n_f-n_a)(r-1)~,
\ee
and $\kappa^\Phi_{GT} = 0$,  to the gauge $(GG)$ and gauge-flavour $(GF)$ CS contact terms (with $\kappa^\Phi_I=\kappa_{GG}$, in the conventions of section~\ref{subsec:bareCS}). We then have:
\be
K= k+ \half(n_f+n_a)~, \qquad L=l~,
\ee
for the bare CS levels for the gauge symmetry. We also choose the bare levels:
\bea
&K_{GA}={\begin{cases}  \Theta(-k) (n_f-n_a) \qquad\quad &\text{if}\quad|k|\geq \half |n_f-n_a|~,\\
\sign(n_f-n_a) (\half |n_f-n_a|- k) \qquad &\text{if}\quad  |k| < \half |n_f-n_a|~,
\end{cases}  }\\
 &K_{GR}= K_{GA} (r-1)~, 
\eea
 with $\Theta(x)$ the Heavyside step function,%
\footnote{Here defined as $\Theta(x)=1$ if $x>0$ and $\Theta(x)=0$ if $x\leq 0$.} 
and:
\be
K_{GT}=1~,
\ee
which corresponds to a standard FI term. Finally, we take all the  flavour bare CS levels to vanish:
\be
K_{SU(n_f)}=K_{SU(n_a)}=K_{AA}=K_{TT}=K_{AT}= K_{RA}=K_{RT}=K_{RR}=K_g=0~.
\ee
We then have $\kappa=\kappa^\Phi$ for the flavour symmetry contact terms in the UV, with:
\bea\label{kappaF SQCD}
&\kappa_{SU(n_f)}^\Phi =  -\half N_c~, &&\kappa_{SU(n_a)}^\Phi = -\half N_c~,\\
& \kappa_{AA}^\Phi= - \half (n_f+n_a)N_c~, \quad && \kappa_{TT}^\Phi= 0~, \\
& \kappa_{RA}^\Phi= - \half (n_f+n_a)N_c (r-1)~, \quad && \kappa_{RT}^\Phi= 0~,   \\
& \kappa_{RR}^\Phi=- \half (n_f+n_a)N_c (r-1)^2+ \half N_c^2~, \qquad && \kappa_{AT}^\Phi= 0~,\\
& \kappa^\Phi_g= - \half (n_f+n_a)N_c + N_c^2~.
\eea
Note that the parameterisation of the R-symmetry through the arbitrary $R$-charge $r=R[Q_i]=R[\t Q^j]$  is somewhat redundant, since one can always mix $U(1)_R$ with the axial symmetry $U(1)_A$, as $R \rightarrow R + \Delta r \, Q_A$. By considering the minimal coupling to background gauge fields, one finds that a shift $r\rightarrow r + \Delta r$ leads to the following shifts of the bare CS levels:
\bea
&	K_{RA} \rightarrow K_{RA} +\Delta r\, K_{AA}~,\quad &&  K_{RI} \rightarrow K_{RI} +\Delta r\, K_{AI}\, \; (I \neq A)~, \\
&	K_{RR} \rightarrow K_{RR} + 2\Delta r\, K_{RA} + \left(\Delta r\right)^2 K_{AA}~,\qquad 
\eea
 and similarly for the CS contact terms themselves.

\medskip
\noindent
{\bf Bethe equations for SQCD.} Let us introduce the flavour parameters:
\be
y_i \;\;  (i=1, \cdots, n_f)~, \quad \prod_{i=1}^{n_f} y_i=1~,\qquad\qquad
\t y_j\;\;  (j=1, \cdots, n_a)~, \quad \prod_{j=1}^{n_a}\t  y_j=1~,
\ee
for $SU(n_f)\times SU(n_a)$,  as well as $y_A=e^{2\pi i \nu_A}$ for $U(1)_A$ and $q=e^{2\pi i \tau}$ for $U(1)_T$. 
The Bethe equations \eqref{BVE 0} for SQCD$[N_c, k, l, n_f, n_a]$ can be written as:
\be
\prod_{i=1}^{n_f} (y_i - x_a y_A)- (-1)^{l+k + \half(n_f+n_a)} q\, y_A^{K_{GA}} \prod_{j=1}^{n_a} (x_a - \t y_j y_A)\,\, x_a^{k+\half (n_f-n_a)} \, \left( \prod_{b=1}^{N_c} x_b \right)^l=0~,
\ee
for $a=1, \cdots, N_c$.
For $l=0$,  we have $N_c$ decoupled equations, but in general we have a coupled system of $N_c$ equations in $N_c$ variables $x_a$. Then, the Gr\"obner basis techniques described in section~\ref{subsec:Grob tech} become particularly useful.

\subsection{Unitary SQCD and its dual description}\label{subsec:duality summary}
The simplest observable of unitary SQCD is the flavored Witten index,
\be
{\bf I}_W(N_c, k, l, n_f, n_a) \equiv  Z_{T^2\times S^1}[\text{SQCD}[N_c, k, l, n_f, n_a]]~.
\ee
For $l=0$, we have \cite{Closset:2016arn}:
\be
{\bf I}_W(N_c, k, 0,  n_f, n_a) = \mat{N_c^D + N_c \\ N_c}~,
\ee
where we defined the `dual rank':
\be\label{NcD def}
N_c^D\equiv \begin{cases} \half(n_f+n_a)+|k|- N_c \qquad\quad &\text{if}\quad|k|\geq | k_c |~,\\
\max(n_f, n_a)- N_c \qquad &\text{if}\quad  |k|\leq |k_c|~,
\end{cases}
\ee
with the `chirality' parameter:%
\footnote{Using some slight abuse of terminology and following~\protect\cite{Benini:2011mf}, we call the 3d SQCD theory `chiral' if $n_f\neq n_a$. This is because the 4d analogue of that theory would be chiral in the usual sense. (One similarly talks about the 3d $\CN=2$ chiral multiplet, it being the dimensional reduction of the 4d $\CN=1$ chiral multiplet.)}
\be\label{def kc}
 k_c \equiv {n_f-n_a\ov 2}~.
\ee
Unitary SQCD with $l=0$ with gauge group $U(N_c)_k$ has an infrared-dual description in terms of a $U(N_c^D)_{-k}$ gauge group, with a particular matter content that depend on the parameters $k$, $n_f$ and $n_a$. There are  four distinct cases to consider  \cite{Benini:2011mf}:
\begin{itemize}
\item[(i)] {\bf Aharony dual.} For $k=0$, $n_f=n_a\equiv N_f$, we have the $U(N_f-N_c)_0$ dual description, known as the Aharony dual~\cite{Aharony:1997bx}. The dual description involves $N_f^2$ gauge-invariant chiral multiplets (the `mesons' of the `electric' SQCD description), as well as two  additional singlets charged under the topological symmetry (the `monopoles' of the electric description).

  \item[(ii)]  {\bf Minimally chiral case.} For $k\neq 0$ and $|k|> |k_c|$, we have a $U(N_c^D)_{-k}$ description with $n_f n_a$ mesons and no monopole singlets. We call these the `minimally chiral' theories. In the non-chiral case,  $n_f=n_a\equiv N_f$, we have a $U(N_f+|k|-N_c)_{-k}$ dual gauge group, and this is known as the Giveon-Kutasov duality~\cite{Giveon:2008zn}.
  
    \item[(iii)]   {\bf Marginally chiral case.} For $k\neq 0$ and $|k|=|k_c|$, we have a $U(N_c^D)_{-k}$ description with $n_f n_a$ mesons and one monopole singlet. We call these the `marginally chiral' theories.
  
    \item[(iv)]  {\bf Maximally chiral case.}  For $|k|< |k_c|$, we have a $U(n_f-N_c)_{-k}$ or $U(n_a-N_c)_{-k}$ description if $n_f >n_a$ or $n_a >n_f$, respectively, with $n_f n_a$ mesons and no monopole singlets. We call these the `maximally chiral' theories.
  
\end{itemize}

\medskip
\noindent
{\bf Unitary SQCD for general $l$.}
 In the general case with arbitrary $l\in \Z$, the computation of the Witten index becomes more involved, as we will discuss momentarily. The theory also has a very interesting `magnetic' dual description with a product unitary gauge group \cite{Nii:2020ikd,Amariti:2021snj}:
\be
G^D= U(N_c^D)\times U(1)~,
\ee
with $N_c^D$ defined  in \eqref{NcD def}, if $|k|\geq |k_c|$ (if $|k|<|k_c|$, the dual gauge group remains $U(N_c^D)$). There are again four cases to consider, as for $l=0$. We will study these dualities in more details in section~\ref{sec:SQCDlnon0}. Using the companion matrix method, we verified that the twisted indices, $Z_{\Sigma_g\times S^1}$, match exactly across all these dualities.%
\footnote{At least in many examples, when running the algorithm on a laptop computer.}
 Here, let us only summarise their key features:

\begin{itemize}
\item[(i)] {\bf Amariti-Rota dual.} For $k=0$, $n_f=n_a\equiv N_f$, we have a dual gauge theory:
\be\label{AR dual gauge group}
U(N_f- N_c\underbracket[0.7pt][7pt]{)_{0,\, 0} \times U(1}_0)_{l}~,
\ee
which was first derived in \cite{Amariti:2021snj}. The line connecting the gauge factors  denotes a (vanishing) mixed CS level,  $k_{12}=0$. Importantly,  the theory also has matter fields charged under both gauge factors. We will discuss this duality in more detail in section~\ref{subsec:AR dual}.

  \item[(ii)]  {\bf Minimally chiral case.} For $k\neq 0$ and $|k|> |k_c|$, we have a dual gauge theory:
  \be
U\big(|k|+\half(n_f+n_a)-N_c \underbracket[0.7pt][7pt]{\big)_{-k, \, -k +\sign(k) N_c} \times U(1}_{\sign(k)})_{l+\sign(k)}~,
  \ee 
  and there is no matter charged under the $U(1)_{l\pm 1}$ factor. It only couples to the $U(N_c^D)$ sector through the mixed CS level $k_{12}= \sign(k)=\pm 1$.
    
    \item[(iii)]   {\bf Marginally chiral.} For $k\neq 0$ and $|k|= |k_c|$, we have a dual description:
      \be
U( \max(n_f, n_a)- N_c\underbracket[0.7pt][7pt]{)_{-k, \, -k +\half \sign(k) N_c} \times U(1}_{\half \sign(k)})_{l+\half \sign(k)}~,
  \ee 
    and there is some matter charged under both gauge groups. 
  
    \item[(iv)]  {\bf Maximally chiral case.}  For $|k|< |k_c|$, we have a dual gauge theory:
      \be
U( \max(n_f, n_a)- N_c)_{-k,\, -k +l N_c^D}~,
  \ee 
   similarly to the $l=0$ case.
\end{itemize}
When $l=0$, these dualities can be reduced to the previous unitary dualities.
 The knowledge of the dual description immediately yields some non-trivial information. For instance, whenever $N_c^D <0$ the Witten index vanishes and therefore supersymmetry could be broken.  The limiting cases of the dualities for which $N_c^D=0$ also give us `$s$-confining' phases (an IR description in terms of chiral multiplets only) if $l=0$ or if $|k|< |k_c|$.

\subsection{Twisted index for $U(N_c)_{0, l N_c}$, $N_f$ SQCD}

Consider the $U(N_c)_{0,\, l N_c}$ gauge theory with $N_f$ pairs of fundamental and antifundamental matter. We find that the Witten index of this theory is given by:
\be\label{IW 0lNc}
{\bf I}_W(N_c,0,l,N_f, N_f) =  {N_f + |l| N_c\ov N_f} \mat{N_f \\N_c}~.
\ee
For $l=0$, this is a well-known result which follows from the fact that the vacua are in one-to-one correspondence with the sets of $N_c$ distinct roots of a degree-$N_f$ polynomial \cite{Closset:2016arn}. 
 For $l\neq 0$, we proceed as follows. By considering certain limits in the space of flavor fugacities, as done in \cite{Closset:2017bse} in a similar context, one can show that the number of Bethe vacua satisfies the recursion relation:
 \be\label{recursion NcNf}
 {\bf I}_W(N_c, 0,l, N_f,N_f) = {\bf I}_W(N_c, 0,l, N_f-1,N_f-1)+ {\bf I}_W(N_c-1, 0,l, N_f-1,N_f-1)~.
 \ee
Furthermore, one can show that:
 \be
  {\bf I}_W(1, 0, l, N_f, N_f) =N_f+ |l|~, \qquad \qquad  {\bf I}_W(N_f, 0, l, N_f, N_f) = 1+ |l|~.
 \ee
 The first equality follows from the fact that, for $N_c=1$, we have an abelian gauge theory $U(1)_l$ with $N_f$ pairs of electrons of charge $\pm 1$, whose index was computed in~\cite{Intriligator:2013lca}. The second equality follows from the fact that, for $N_c=N_f$, the  Amariti-Rota dual \eqref{AR dual gauge group} is a $U(1)_{l}$ theory with one flavour pair. With these initial conditions, one can solve the recursion relation \eqref{recursion NcNf} to obtain \eqref{IW 0lNc}.

\medskip
\noindent
{\bf Twisted indices: a few examples.} Let us now consider a few simple examples of twisted indices, computed using the companion matrix method. First, consider the $U(2)_0$ theory ($l=0$) with $N_f=2$ flavours and $r=0$. There is a single Bethe vacuum and the index is given by:
\be\label{Z4m}
Z_{\Sigma_g\times S^1}^{U(2)_0, \, N_f=2}(y_A, \chi, \t \chi) = \left(1- \chi \t\chi y_A^2 + (\chi^2+ \t\chi^2-2) y_A^4- \chi\t\chi y_A^6 + y_A^8\right)^{g-1} \equiv Z_{4 M}~.
\ee
This theory is dual to four free chiral multiplets, the mesons ${M_i}^j = Q_i \t Q^j$, and this is apparent in the index. Here we introduced the characters $\chi= y_1+ y_2$ and $\t\chi= \t y_1 + \t y_2$  for the $SU(2)\times SU(2)$ flavour symmetry. 

As another example, let us consider the abelian theory $U(1)_{l}$ with CS level $l$ and $N_f=1$ flavour. This theory has $|l|+1$ Bethe vacua. On the sphere ($g=0$), we have:
\be
Z_{S^2 \times S^1}^{U(1)_l , \, N_f=1}(y_A, q) ={1\ov 1-y_A^2}~,
\ee
for any $l$. At genus $g=1$, we have the Witten index, $Z_{T^3}=|l|+1$. At genus $g=2$, we find:
\be
Z_{\Sigma_2 \times S^1}^{U(1)_l , \, N_f=1}(y_A, q) =\begin{cases}1-2l -l^2 - (l-1)^2 y_A^2 - \delta_{l, -1} \,\left(q+q^{-1}\right)\qquad &\text{if} \;\; l<0 \\
(1+l)^2+ (l^2-2l-1) y_A^2+ \delta_{l, 1}\,\left(q+q^{-1}\right) y_A^2  &\text{if}\;\; l\geq 0~,
 \end{cases}
\ee
and similar formulas can be worked out for any $g>2$. In particular, we observe that the index terminates (thus there is a finite number of states on any $\Sigma_g$ with $g>0$), and that states charged under the topological symmetry (weighted by $q$) only appear for  $0<|l|\leq g-1$.

Next, consider $U(2)_{0, 2l}$ with $N_f=2$ and $r=0$, a theory with $|l|+1$ Bethe vacua.  By explicit computation, one can check that the index factorises as:
\be
Z_{\Sigma_g\times S^1}^{U(2)_{0, 2l}, \, N_f=2}(y_A, \chi, \t \chi) =(1-y_A^4)^{1-g} \; Z_{4 M}  \; Z_{\Sigma_g\times S^1}^{U(1)_l, \, N_f=1}(y_A^2, q)~,
\ee
with $Z_{4 M}$ as in \eqref{Z4m}. This can be precisely explained in terms of the Amariti-Rota duality, to be discussed in detail in section~\ref{subsec:AR dual}. In particular, the prefactor $(1-y_A^4)^{1-g}$ should be written as $(1-y_A^{-4})^{1-g} \times y_A^{-4(g-1)}\times (-1)^{g-1}$, corresponding to a chiral multiplet of $U(1)_A$ charge $-4$ and to some flavour CS levels $K_{RA}=-4$ and $K_{RR}=1$.

\subsection{The Witten index for $n_f=n_a$ and $k\neq 0$}

Consider the theory with $n_f=n_a\equiv N_f$ flavours and $k\neq 0$. Consider first the case $N_f=0$, which is the $\CN=2$ supersymmetric Chern-Simons theory $U(N_c)_{k,\, k+l N_c}$. We claim that the Witten index of this theory is given by:
\be\label{IW Nf0}
 {\bf I}_W(N_c, k, l, 0,0) = {|k+  l N_c|\ov |k|} \mat{|k|\\ N_c}~,
\ee
for $|k| \geq N_c$, with the understanding that the index vanishes for $k< N_c$ (supersymmetry is broken in that case). This can be understood from the fact that:
\be\label{Z_N quotient UNkl}
U(N_c)_{k, \, k+l N_c} \cong {SU(N_c)_k\times U(1)_{N_c(k + l N_c)}\ov \Z_{N_c}}~.
\ee
Setting $x_a= \t x_a x_0$ with the constraint $\prod_{a=1}^{N_c}\t x_a=1$,  the Bethe equations for the pure 3d $\CN=2$ CS theory read:
\be\label{UNc CS bethe}
\t x_a^k  = (-1)^{k+l}q^{-1} x_0^{-(k+ l N_c)}~,\qquad a=1, \cdots, N_c~,
\ee
which implies that $x_0$ is proportional to a $N_c(k+l N_c)$-th root of unity. Plugging back this solution into \eqref{UNc CS bethe}, we have the Bethe equations for the $SU(N_c)_k$ supersymmetric CS theory. Taking into account the redundancy in our parameterisation $x_a= \t x_a x_0$ and the overall $\Z_{N_c}$ quotient in \eqref{Z_N quotient UNkl}, one obtains  \eqref{IW Nf0} written as:
\be
 {\bf I}_W(N_c, k, l, 0,0) =  {|k+l N_c|\ov N_c} \times {\bf I}_W[SU(N_c)_k]~, \quad\qquad {\bf I}_W[SU(N_c)_k]= \mat{|k|-1\\ N_c-1}~,
\ee
using the $SU(N_c)_k$ Witten index  computed in \cite{Witten:1999ds, Ohta:1999iv}.

\medskip
\noindent
For any $N_f >0$, we have the recursion relation:
\be\label{recursion Nc Nf k l}
 {\bf I}_W(N_c, k, l,  N_f, N_f) = {\bf I}_W(N_c, k, l,N_f-1, N_f-1)+ {\bf I}_W(N_c-1, k, l,N_f-1, N_f-1)~,
 \ee
 similarly to \eqref{recursion NcNf}. 
Thus, given the result \eqref{IW Nf0}, we can compute the Witten index recursively. For instance, one easily finds:
\be\label{IW Nc Nf klpos}
 {\bf I}_W(N_c, k, l,  N_f,N_f) = {N_f+ k + l N_c \ov  N_f+k} \mat{ N_f+k \\ N_c} \quad\; \text{if} \; k> 0 \; \text{and}\;  l\geq 0~,
\ee
where we also assumed that $N_c \leq  N_f +k$.
For general values of $k$ and $l$, the recursive definition \eqref{recursion Nc Nf k l} (with the boundary condition \eqref{IW Nf0}) gives us the explicit formula: 
\be
 {\bf I}_W(N_c, k, l,  N_f,N_f) =\sum_{j=0}^{N_f} {|k+ l(N_c-j)|\ov |k|}\, \mat{N_f\\ j}\mat{|k|\\ N_c-j}~.
\ee
 We checked numerically, in a large number of examples, that the index so obtained matches the number of Bethe vacua computed by Gr\"obner basis methods. 
 
\begin{table}[t]
\renewcommand{\arraystretch}{1.1}
\centering
\be\nn
 \begin{array}{|c||c|c|c|c|c|c|c|c|c|c|c|c|c|c|c|c|c|c|c|c|c|}
 \hline
 k \backslash l & -10 & -9 & -8 & -7 & -6 & -5 & -4 & -3 & -2 & -1 & 0 & 1 & 2 & 3 & 4 & 5
   & 6 & 7 & 8 & 9 & 10 \\ \hline \hline
 0 & 15 & 14 & 13 & 12 & 11 & 10 & 9 & 8 & 7 & {\bf 6} & {\bf 6} & {\bf 6} & 7 & 8 & 9 & 10 & 11 &
   12 & 13 & 14 & 15 \\ \hline
 1 & 23 & 21 & 19 & 17 & 15 & 13 & 11 & 9 & 7 & {\bf 6} & {\bf 6} & {\bf 6} & 7 & 9 & 11 & 13 & 15
   & 17 & 19 & 21 & 23 \\ \hline
 2 & 30 & 27 & 24 & 21 & 18 & 15 & 12 & 9 & {\bf 6} & {\bf 6} & {\bf 6} & 9 & 12 & 15 & 18 & 21 &
   24 & 27 & 30 & 33 & 36 \\ \hline
 3 & 36 & 32 & 28 & 24 & 20 & 16 & 12 & 8 &{\bf 6}&{\bf 6}& 10 & 14 & 18 & 22 & 26 & 30
   & 34 & 38 & 42 & 46 & 50 \\ \hline
 4 & 41 & 36 & 31 & 26 & 21 & 16 & 11 &{\bf 6}&{\bf 6}& 10 & 15 & 20 & 25 & 30 & 35 & 40
   & 45 & 50 & 55 & 60 & 65 \\ \hline
 5 & 45 & 39 & 33 & 27 & 21 & 15 & 9 &{\bf 6}& 10 & 15 & 21 & 27 & 33 & 39 & 45 & 51
   & 57 & 63 & 69 & 75 & 81 \\ \hline
 6& 48 & 41 & 34 & 27 & 20 & 13 &{\bf 6}& 10 & 14 & 21 & 28 & 35 & 42 & 49 & 56 &
   63 & 70 & 77 & 84 & 91 & 98 \\ \hline
 7 & 50 & 42 & 34 & 26 & 18 & 10 & 10 & 14 & 20 & 28 & 36 & 44 & 52 & 60 & 68 &
   76 & 84 & 92 & 100 & 108 & 116 \\ \hline
 8 & 51 & 42 & 33 & 24 & 15 & 10 & 14 & 18 & 27 & 36 & 45 & 54 & 63 & 72 & 81 &
   90 & 99 & 108 & 117 & 126 & 135 \\ \hline
 9 & 51 & 41 & 31 & 21 & 15 & 14 & 18 & 25 & 35 & 45 & 55 & 65 & 75 & 85 & 95 &
   105 & 115 & 125 & 135 & 145 & 155 \\ \hline
 10 & 50 & 39 & 28 & 21 & 14 & 18 & 22 & 33 & 44 & 55 & 66 & 77 & 88 & 99 & 110
   & 121 & 132 & 143 & 154 & 165 & 176 \\ \hline
\end{array}
\ee
\caption{Witten index for $U(2)_{k, k+2l}$ with $n_f=4$ fundamentals, for some values of $k, l$. The values with $k<0$ can be obtained by parity (that is, the index for some $(k,l)$ is the same as the index for $(-k,-l)$). The cases with the `geometric value' $ {\bf I}_W=6$ are given in bold. }
\label{tab: Gr24}
\end{table}
 \subsection{Twisted index for chiral theories}
Finally, let us briefly discuss the case of chiral theories.   For any SQCD theory with $n_f n_a >0$, we have a recursion relation:
\be\label{recursion Nc nf na k l}
 {\bf I}_W(N_c, k, l,  n_f, n_a) = {\bf I}_W(N_c, k, l,n_f-1, n_a-1)+ {\bf I}_W(N_c-1, k, l,n_f-1, n_a-1)~.
 \ee
Assuming that $n_f >n_a$ without loss of generality, the question is then to find the Witten index for $n_a=0$,
\be\label{IW na0}
 {\bf I}_W(N_c, k, l,  n_f, 0)~.
\ee
The computation of the index  \eqref{IW na0}  for general $k$ and $l$ is part of a rather rich story, which will be addressed more thoroughly in future work~\cite{toappear2023}. Here, let us just mention that we can easily compute it using Gr\"obner basis methods, at least for small enough values of the parameters $N_c, n_f$ and $k, l$. Some examples are displayed in tables~\ref{tab: Gr24} and~\ref{tab: Gr25}, where we computed the index for $U(N_c)_{k, k+lN_c}$ with $n_f$ fundamentals for $(N_c, n_f)=(2,4)$ and $(2,5)$.
\begin{table}[t]
\renewcommand{\arraystretch}{1.1}
\centering
\be\nn
 \begin{array}{|c||c|c|c|c|c|c|c|c|c|c|c|c|c|c|c|c|c|c|c|c|c|}
 \hline
 k \backslash l & -10 & -9 & -8 & -7 & -6 & -5 & -4 & -3 & -2 & -1 & 0 & 1 & 2 & 3 & 4 & 5
   & 6 & 7 & 8 & 9 & 10 \\ \hline \hline
 {1\ov 2} & 27 & 25 & 23 & 21 & 19 & 17 & 15 & 13 & 11 &{\bf 10}&{\bf 10}&{\bf 10}& 11 & 12 & 13 &
   14 & 15 & 17 & 19 & 21 & 23 \\ \hline
 {3\ov 2} & 34 & 31 & 28 & 25 & 22 & 19 & 16 & 13 &{\bf 10}&{\bf 10}&{\bf 10}&{\bf 10}& 12 & 15 & 18 &
   21 & 24 & 27 & 30 & 33 & 36 \\ \hline
 {5\ov 2} & 40 & 36 & 32 & 28 & 24 & 20 & 16 & 12 &{\bf 10}&{\bf 10}&{\bf 10}& 14 & 18 & 22 & 26 &
   30 & 34 & 38 & 42 & 46 & 50 \\ \hline
 {7\ov 2} & 45 & 40 & 35 & 30 & 25 & 20 & 15 &{\bf 10}&{\bf 10}&{\bf 10}& 15 & 20 & 25 & 30 & 35 &
   40 & 45 & 50 & 55 & 60 & 65 \\ \hline
 {9\ov 2} & 49 & 43 & 37 & 31 & 25 & 19 & 13 &{\bf 10}&{\bf 10}& 15 & 21 & 27 & 33 & 39 & 45 &
   51 & 57 & 63 & 69 & 75 & 81 \\ \hline
 {11\ov 2} & 52 & 45 & 38 & 31 & 24 & 17 &{\bf 10}&{\bf 10}& 15 & 21 & 28 & 35 & 42 & 49 & 56 &
   63 & 70 & 77 & 84 & 91 & 98 \\ \hline
 {13\ov 2} & 54 & 46 & 38 & 30 & 22 & 14 &{\bf 10}& 15 & 20 & 28 & 36 & 44 & 52 & 60 & 68 &
   76 & 84 & 92 & 100 & 108 & 116 \\ \hline
 {15\ov 2} & 55 & 46 & 37 & 28 & 19 &{\bf 10}& 15 & 20 & 27 & 36 & 45 & 54 & 63 & 72 & 81 &
   90 & 99 & 108 & 117 & 126 & 135 \\ \hline
 {17\ov 2} & 55 & 45 & 35 & 25 & 15 & 15 & 20 & 25 & 35 & 45 & 55 & 65 & 75 & 85 & 95 &
   105 & 115 & 125 & 135 & 145 & 155 \\ \hline
 {19\ov 2} & 54 & 43 & 32 & 21 & 15 & 20 & 25 & 33 & 44 & 55 & 66 & 77 & 88 & 99 & 110 &
   121 & 132 & 143 & 154 & 165 & 176 \\ \hline
 {21\ov 2} & 52 & 40 & 28 & 21 & 20 & 25 & 30 & 42 & 54 & 66 & 78 & 90 & 102 & 114 &
   126 & 138 & 150 & 162 & 174 & 186 & 198 \\ \hline
\end{array}
\ee
\caption{Witten index for $U(2)_{k, k+2l}$ with $n_f=5$ fundamentals, for some values of $k, l$.  The cases with the `geometric value' $ {\bf I}_W=10$ are given in bold.}
\label{tab: Gr25}
\end{table}
Let us comment on the fact that, for some values of the parameter, this index takes the `geometric value' which saturates the bound:
\be
 {\bf I}_W(N_c, k, l,  n_f, 0)\, \geq\,  {\bf I}_W^{\rm Higgs} = \chi({\rm Gr}(N_c, n_f))= \mat{n_f\\ N_c}~.
\ee
This lowest value of the index, $ {\bf I}_W^{\rm Higgs}$, is the contribution from the Higgs branch of the gauge theory in a phase with vanishing masses and a positive FI parameter. That Higgs branch is the Grassmannian ${\rm Gr}(N_c, n_f)$, and the Witten index of this geometric phase is given by its Euler characteristic. For general values of $k, l$, there are also a number of additional contributions to the index from  topological and mixed topological-geometric vacua, similarly to the abelian cases considered in~\cite{Intriligator:2013lca}.

\medskip
\noindent
{\bf More examples of twisted indices: Grassmannian theories.}. Let us consider the genus-$0$ index for the ${\rm Gr}(2,4)$ `Grassmannian theories', defined as the $U(2)_{k, k+2l}$, $n_f=4$ theories with CS levels such that ${\bf I}_W=6$. There are 15 such theories. Setting the $SU(4)$ flavour fugacities to $y_i=1$ for simplicity, the $S^2\times S^1$ twisted index of these theories is given by:
\be
Z_{S^2\times S^1} =\begin{cases}
0 \quad &\text{if}\; (k,l)\in \{(0,-1),(0,0),(0,1),(1,-1),(1,0),(1,1),(5,-4)\}~,\\
1 &\text{if}\; (k,l)\in \{(2,-1),(2,0),(3,-2),(3,-1),(4,-2),(5,-3) \}~,\\
{1\ov 1-q^2} &\text{if}\; (k,l)\in \{(2,-2),(4,-3) \}~.
\end{cases}
\ee
We can similarly compute the higher-genus index. For instance, for this ${\rm Gr}(2,4)$ theory with $(k,l)=(2,-1)$ (and $y_i=1$), we find:
\be\nn
Z_{\Sigma_g\times S^1}^{U(2)_{2,0}, \, n_f=4}=
\begin{cases}
 1  & \text{for} \; g=0~, \\
 6  & \text{for} \; g=1~, \\
 24 q-30 q^2+2 q^3 & \text{for} \; g=2~, \\
 288 q^2-1256 q^3+1188 q^4-216 q^5+4 q^6& \text{for} \; g=3~, \\
 1920 q^3-672 q^4+31896 q^5-28944 q^6+5832 q^7+24 q^8-40 q^9 & \text{for} \; g=4~, \\
 16896 q^4+428032 q^5-157280 q^6-1098432 q^7+975760 q^8 &\\
 -243968 q^9+10464 q^{10}+1984 q^{11}-112 q^{12} & \text{for} \; g=5~, \\
  \end{cases}
   \ee
for the first few values of $g$. It would be very interesting to understand these and many similar results for the twisted index from an explicit quantisation of the theory on $\Sigma_g$ with the $A$-twist, similarly to \cite{Bullimore:2018yyb, Bullimore:2020nhv}.

\section{Infrared dualities for $U(N_c)_k$ SQCD, revisited}\label{sec:SQCDl0}

In this section and the next, we revisit and clarify some aspects of the infrared dualities for 3d $\CN=2$ supersymmetric SQCD. We give a complete definition of the  `magnetic' dual theory in all cases, including various Chern-Simons contact terms for the flavour symmetry, in the gauge-invariant conventions of sections~\ref{subsec:bareCS} and~\ref{subsec:sqcd flavour def}.

\subsection{Aharony duality ($k=l=0$, $n_f=n_a$)}\label{subsec:Ah dual}
If we set $k=l=0$ and $n_f=n_a\equiv N_f$, we have the famous Aharony duality \cite{Aharony:1997gp}:
\be\label{aharony dual}\boxed{
U(N_c)_0~, \; N_f  ({\tiny\yng(1)}\oplus \overline{ {\tiny\yng(1)}})  \qquad
\longleftrightarrow \qquad
U(N_f-N_c)_{0}~, \; N_f  ({\tiny\yng(1)}\oplus \overline{ {\tiny\yng(1)}})~,  \; \left({M_i}^j, \,\mathfrak{T}_+,\,  \mathfrak{T}_-\right)~.}
\ee
This can be viewed as the most fundamental infrared duality for $U(N_c)$ SQCD, in the sense that all the other dualities summarised in section~\ref{subsec:duality summary} can be derived from Aharony duality using appropriate limits~\cite{Benini:2011mf, Amariti:2021snj}. The matter content and the charges of the Aharony dual theory are shown in table~\ref{tab: Aharony duality charges}.
\begin{table}[t]
\renewcommand{\arraystretch}{1.1}
\centering
\be\nn
\begin{array}{|c|c|ccccc|}
\hline
    & U(N_c^D)& SU(N_f) & SU(N_f)  & U(1)_A &  U(1)_T & U(1)_R  \\
\hline
q_j   &{\tiny\yng(1)}&    \bm{1} & \overline{\tiny\yng(1)} & -1   & 0   &1-r \\
\t q^i       &\overline{\tiny\yng(1)}&{\tiny\yng(1)}   & \bm{1}    & -1   & 0   &1-r \\
{M ^j}_i      & \bm{1} &\overline{\tiny\yng(1)}& {\tiny\yng(1)} &  2   & 0   &2 r \\
\T^+       & \bm{1} & \bm{1} & \bm{1} & -N_f   & 1   & r_T\\
\T^-     & \bm{1} & \bm{1} & \bm{1} & -N_f   & -1   & r_T\\
\hline
\end{array}
\ee
\caption{Field content of the Aharony dual theory, with $N_c^D= N_f-N_c$ and $r_T$ given in \protect\eqref{rT def}.}
\label{tab: Aharony duality charges}
\end{table}
We have dual flavours $q_j$ and $\t q^i$ in the fundamental and antifundamental of $U(N_f-N_c)$, respectively.  The gauge singlets ${M_i}^j$ and $\T^\pm$ are identified with the gauge-invariant `mesons' ${M_i}^j= Q_i \t Q^j$ and with the monopole operators, respectively, in the electric description. The monopoles have $R$-charge:
\be\label{rT def}
r_T= - N_f(r-1)- N_c+1~.
\ee
The gauge singlets are coupled to the gauge sector through the superpotential $W=\t q^i {M_i}^j q_j + \T^+ t_+ + \T^- t_-$, where $t_\pm$ are the gauge-invariant monopole operators of the dual theory.

It is clear that we have $K=\half(n_f+n_a)$ and $L=0$ for the $U(N_c^D)$ bare CS levels. Moreover, given our conventions for the electric theory, the magnetic theory must have non-vanishing bare CS levels for the flavour symmetry. We have
\be\label{aharony-bare-cs-levels-1}
	K^{(\text{Ah})}_{SU(N_f)}= N_f-N_c~, \qquad\qquad  \t K^{(\text{Ah})}_{SU(N_f)} = N_f-N_c,
\ee
for the $SU(N_f)\times SU(N_f) $ flavour symmetry and
\bea	\label{Aharony-cs-levels}
& K^{(\text{Ah})}_{TT} &=&\; 1~,\\
&	K^{(\text{Ah})}_{AA} &=&\; 4N_f^2-2N_cN_f~, \\
&	K^{(\text{Ah})}_{AR} &=&\; 2N_f^2 + \left(4N_f^2-2N_cN_f\right)(r-1)~,\\
&	K^{(\text{Ah})}_{RR} &=&\; N_c^2 + N_f^2 + 4N_f^2(r-1) + \left(4N_f^2 - 2N_cN_f\right)(r-1)^2~,\\
&	K^{(\text{Ah})}_g &=&\; 2N_f (N_f-N_c) + 2~.
\eea
for the abelian flavour symmetry (as well as for the gravitational CS contact term)  \cite{Closset:2018ghr}. 
All other flavour and gauge-flavour levels vanish except for
\be
K_{GT}^{(\text{Ah})}= -1~,
\ee
which is the statement that the FI parameter changes sign under the duality -- equivalently, the topological currents of the $U(N_c)$ and $U(N_c^D)$ dual gauge groups are identified up to a sign.

In the limiting case $N_f=N_c$, the dual theory consists of a linear $\sigma$-model with $N_f^2+2$ chiral multiplets ${M^k}_i$, $\T^\pm$ which are coupled through the superpotential
 $W= \T^+ \T^- {\rm det}(M)$.
Finally, for $N_f<N_c$, we either have a quantum-deformed moduli space (for $N_f=N_c-1$) or supersymmetry breaking (for $N_f <N_c-1$)~\cite{Aharony:1997gp}. 

\subsection{Minimally chiral duality with $l=0$}\label{subsec:l0minchir}
Next, let us consider the case $k\neq 0$ with $l=0$ and with the `minimally chiral' condition $|k|>  |k_c|$. 
We have a duality:
\be\label{min chiral dual}
\boxed{
U(N_c)_k~, \;  \left(n_f\, {\tiny\yng(1)}, n_a\, \overline{ {\tiny\yng(1)}}\right)  \qquad
\longleftrightarrow \qquad
U(N_c^D)_{-k}~, \;   \left(n_a\, {\tiny\yng(1)}, n_f\, \overline{ {\tiny\yng(1)}}\right)~,  \; \left({M_i}^j\right)~.}
\ee
with:
\be \label{NcD min chiral}
N_c^D= \half(n_f+n_a)+|k|-N_c~.
\ee
Now the singlet sector only consists of the `mesons' ${M_i}^j $, with the standard  superpotential $W=\t q_i {M_i}^j q_j$ as in 4d Seiberg duality.  The matter content is shown in table~\ref{tab: general dual leq0}. 
\begin{table}[t]
\renewcommand{\arraystretch}{1.1}
\centering
\be\nn
\begin{array}{|c|c|ccccc|c|}
\hline
    & U(N_c^D)& SU(n_f) & SU(n_a)  & U(1)_A &  U(1)_T & U(1)_R & \text{condition} \\
\hline
q_j   &{\tiny\yng(1)}&    \bm{1} & \overline{\tiny\yng(1)} & -1   & 0   &1-r &\\
\t q^i       &\overline{\tiny\yng(1)}&{\tiny\yng(1)}   & \bm{1}    & -1   & 0   &1-r &\\
{M_i}^j      & \bm{1} &\overline{\tiny\yng(1)}& {\tiny\yng(1)} &  2   & 0   &2 r &\\
\hline
\T^+       & \bm{1} & \bm{1} & \bm{1} & -N_f   & 1   & r_T&   k= \half(n_f-n_a)\\
\hline
\T^-     & \bm{1} & \bm{1} & \bm{1} & -N_f   & -1   & r_T&   \; k= -\half(n_f-n_a) \\
\hline
\end{array}
\ee
\caption{Field content of the infrared dual of unitary SQCD with $l=0$. The gauge singlets $\T^\pm$ only appear in the marginally chiral case (or in the Aharony dual), as indicated. Here $N_f\equiv N_c+ N_c^D$.}
\label{tab: general dual leq0}
\end{table}
There are also mixed gauge-flavour bare CS levels:
\be\label{mixed GF min chiral}
K_{GT}=-1~, \qquad K_{GA} = \Theta(k) (n_f-n_a)~, \qquad K_{GR} = \Theta(k) (n_f-n_a)(r-1)~.
\ee
Finally and importantly, we have the bare  flavour CS levels shown in table~\ref{tab: bare levels min chiral}.
 Note that the bare CS levels depend on the sign of $k$, and that changing the sign of $k$ does not simply change the sign of the bare CS levels. This is because of our parity-violating conventions, of course. For any given theory with the `$U(1)_{-\half}$ quantisation' convention, a parity transformation changes the sign of the physical contact terms $\kappa = \kappa^\Phi + K$ as $\kappa \rightarrow -\kappa$ but  the UV `matter' contribution $\kappa^\Phi$ remains the same by convention, as defined in \eqref{kappa matter 1}-\eqref{kappa matter 2}, hence a parity transformation changes the bare CS levels according to:
 \be
 {\rm P} \; : \; K \rightarrow -K - 2 \kappa^\Phi~.
 \ee
Now, if we consider the fact that we chose $K_F=0$ for all the flavour bare CS levels in the `electric' theory, $K_F^{(e)}=0$, irrespective of the sign of $k$, we find that, in the magnetic theory:
\be\label{rel between KF for kpm}
K_F^{(m)}\big|_{k \rightarrow -k} =  - K_F^{(m)} - 2 \kappa_F^{\Phi (m)}+ 2 \kappa_F^{\Phi (e)}~,
\ee
where $\kappa_F^{\Phi (e)}$ and $\kappa_F^{\Phi (m)}$ denote the matter contributions in the electric and magnetic theories, respectively. This gives us the relation between the two columns in table~\ref{tab: bare levels min chiral}.%
\footnote{For instance, consider $K_{AA}$. We have $\kappa_{AA}^{\Phi (e)}= -\half(n_f+n_a)N_c$ according to \eqref{kappaF SQCD}, and $\kappa_{AA}^{\Phi (m)}= -\half(n_f+n_a)N_c - 2 n_f n_a$ in the magnetic theory (including the contribution from the mesons). Then the relation \protect\eqref{rel between KF for kpm} is indeed satisfied by the levels given in table~\protect\ref{tab: bare levels min chiral}.}

\begin{table}
\renewcommand{\arraystretch}{1.2}
\centering
\be\nn
\begin{array}{|c|c|c|}
\hline
    & k > |k_c| & k\leq  - |k_c|\\
\hline\hline
\; K_{SU(n_f)} \;&N_c^D& - N_c + n_a\\
\hline
\; K_{SU(n_a)} \;&N_c^D& - N_c + n_f\\
\hline
\; K_{AA} \;& (n_f+n_a) N_c^D&  4 n_f n_a -(n_f+n_a) N_c\\
\hline
\; K_{TT} \;& -1 & 1\\
\hline
\; K_{AT} \; & 0 & 0\\
\hline
\; K_{RA}\;& (n_f+n_a) N_c^D r &  (n_a+n_f)N_c - 2 n_f n_a+ (4 n_f n_a - (n_f+n_a) N_c)r\\
\hline
\; K_{RT} \;& 0 & 0\\
\hline
\multirow{2}{*}{\; $K_{RR}$ \;}&\multirow{2}{*}{\quad $(-N_c^D + (n_f+n_a)r^2)N_c^D$\quad} &  (N_c-n_f)(N_c-n_a)+2 r ((n_a+n_f)N_c - 2 n_f n_a)\\
   & &  +r^2 (4 n_f n_a -(n_f+n_a) N_c)  \\
\hline
\; K_{g} \;&  (n_f + n_a - 2k)N_c^D-2 &  2 n_f n_a -(n_f+n_a+2 k) N_c +2\\
\hline
\end{array}
\ee
\caption{Flavour bare CS levels for the minimally-chiral $U(N_c^D)_{-k}$ gauge theory ($l=0$), as well as for the marginally chiral case with $k= - |k_c|<0$.}
\label{tab: bare levels min chiral}
\end{table}
This minimally chiral duality, including all the flavour CS levels, can be derived from the Aharony duality by integrating out flavours \cite{Benini:2011mf}, as we review in appendix~\ref{appendix: defs}. 

\subsection{Marginally chiral duality with $l=0$}\label{subsec:l0margchir}
Consider now the `marginally chiral' cases with $|k|= |k_c|$ and $l=0$, with $k_c$ defined in \eqref{def kc}. In this case, the dual gauge theory is coupled to another gauge singlet in addition to the mesons, corresponding to the single gauge-invariant monopole operator $\T^+$ or $\T^-$ in the electric theory (for $k=k_c$ or $k=-k_c$, respectively), as indicated in table~\ref{tab: general dual leq0}.  We then have the duality:
\be\boxed{
U(N_c)_k~, \;  \left(n_f\, {\tiny\yng(1)}, n_a\, \overline{ {\tiny\yng(1)}}\right) \qquad
\longleftrightarrow \qquad
U(N_c^D)_{-k}~, \;   \left(n_a\, {\tiny\yng(1)}, n_f\, \overline{ {\tiny\yng(1)}}\right)~,  \; \left({M_i}^j, \T^\epsilon \right)~,}
\ee
with a dual superpotential $W=\t q^i {M_i}^j q_j + \T^\epsilon t_\epsilon$ for $\epsilon =\pm$. Note that:
\be
N_c^D= \half(n_f+n_a)+|k|- N_c= \max(n_f, n_a)- N_c,
\ee
in this case. Note also that the $R$-charge of $\T^\epsilon$ is given by:
\be
r_T= -(N_c^D + N_c) (r - 1) - N_c + 1~.
\ee
 The mixed gauge-flavour CS levels are the same as in \eqref{mixed GF min chiral}. The flavour bare CS levels have to be carefully determined by real-mass deformation from Aharony duality, like for the minimally chiral case (see appendix~\ref{appendix: defs}). For $k=|k_c|>0$, those levels are given in table~\ref{tab: bare levels marg chiral l0}, while for $k=-|k_c|$ they were given in table~\ref{tab: bare levels min chiral}. 
\begin{table}
\renewcommand{\arraystretch}{1.2}
\centering
\be\nn
\begin{array}{|c|c|c|}
\hline
    & k = |k_c|~, \;\; n_f >n_a & k = |k_c|~, \;\; n_a >n_f \\
\hline\hline
\; K_{SU(n_f)} \;&N_c^D& N_c^D\\
\hline
\; K_{SU(n_a)} \;&N_c^D& N_c^D\\
\hline
\; K_{AA} \;&n_f^2 +(n_f+n_a)N_c^D & n_a^2+ (n_f+n_a)N_c^D \\
\hline
\; K_{TT} \;&0 & 0\\
\hline
\; K_{AT} \; &- n_f & n_a\\
\hline
\; K_{RA}\;&\;\;   -N_c^D n_f + r( n_f^2 +(n_f+n_a) N_c^D)\;\; &\;\;    -N_c^D n_a + r(n_a^2 +(n_f+n_a) N_c^D)\;\; \\
\hline
\; K_{RT} \;& N_c^D-  r n_f &-N_c^D+ r n_a  \\
\hline
\multirow{2}{*}{\; $K_{RR}$ \;}&  - 2 r N_c^D n_f  & -2 r N_c^D n_a    \\
 &+r^2 (n_f^2 +(n_f+n_a)N_c^D ) &+r^2( n_a^2+ (n_f+n_a)N_c^D) \\
\hline
\; K_{g} \;& 2 n_a  N_c^D & 2 n_f  N_c^D\\
\hline
\end{array}
\ee
\caption{Flavour bare CS levels for the marginally chiral dual with $k= |k_c|>0$ ($l=0$).}
\label{tab: bare levels marg chiral l0}
\end{table}

\subsection{Maximally chiral duality with $l=0$}\label{subsec:l0maxgchir}
Finally, we have the maximally chiral case when  $|k|< |k_c|$ and $l=0$.  The general form of the duality is the same as in for minimally chiral case \eqref{min chiral dual}, with the matter content of table~\ref{tab: general dual leq0}, but with the dual rank given by:
\be
N_c^D= \max(n_f, n_a)- N_c~,
\ee
and with the mixed gauge-flavour CS levels:
\be\label{mixed GF max chiral}
K_{GT}=-1~, \qquad K_{GA} =\sign(k_c) (k+|k_c|)~, \qquad K_{GR} = \sign(k_c) (k+|k_c|)(r-1)~.
\ee
Finally, we have the flavour bare CS levels given in table~\ref{tab: bare levels max chiral}. 
\begin{table}
\renewcommand{\arraystretch}{1.2}
\centering
\be\nn
\begin{array}{|c|c|c|}
\hline
    & |k| < |k_c|~,\;\; n_f >n_a & |k| < |k_c|~,\;\; n_a >n_f \\
\hline\hline
\; K_{SU(n_f)} \;& k +\half(n_f+n_a)-N_c& n_a-N_c\\
\hline
\; K_{SU(n_a)} \;&n_f-N_c&  k +\half(n_f+n_a)-N_c\\
\hline
 \; K_{AA}\;& K_{AA}^+\equiv  n_f^2+  3 n_f(k-k_c) + 2 N_c^D(n_f-k_c) &K_{AA}^- \equiv  n_a^2+ 3 n_a(k+k_c) + 2 N_c^D(n_a+k_c) \\
\hline
\; K_{TT} \;&0 & 0\\
\hline
\; K_{AT} \; &- n_f & n_a\\
\hline
\multirow{2}{*}{\;$K_{RA}$\;}&\;\; K_{RA}^{(0)+}+ r K_{AA}^+~,  \;\; &\;\;  K_{RA}^{(0)-}+ r K_{AA}^-~,\;\; \\
&K_{RA}^{(0)+} \equiv  -  (k-k_c)(n_f+ N_c^D)+ n_f N_c^D & K_{RA}^{(0)-} \equiv -  (k+k_c)(n_a+ N_c^D)+ n_a N_c^D \\
\hline
\; K_{RT} \;& N_c^D- r n_f&-N_c^D+r n_a  \\
\hline
\; K_{RR} \;& N_c^D (k-k_c) + 2 r K_{RA}^{(0)+} +r^2 K_{AA}^+ &  N_c^D (k+k_c)  + 2 r K_{RA}^{(0)-} +r^2 K_{AA}^-  \\
\hline
\; K_{g} \;& 2 n_a  N_c^D & 2 n_f  N_c^D\\
\hline
\end{array}
\ee
\caption{Flavour bare CS levels for the maximally chiral dual, $|k|< |k_c|$ ($l=0$).}
\label{tab: bare levels max chiral}
\end{table}

One can check the matching of partition functions across  these dualities explicitly. Given the precise definition  of the dual theories, including all the bare CS levels, one can apply the formalism of section~\ref{sec:twisted indices} to verify that the twisted indices of dual theories exactly agree:
\be\label{Zg dual eq}
Z_{\Sigma_g\times S^1}[\text{SQCD}] =Z_{\Sigma_g \times S^1}[\text{dual SQCD}]~.
\ee
 The proof of this equality for  Aharony duality  in the `$U(1)_{-\half}$ quantisation'  was given in~\cite{Closset:2018ghr}, building on previous works~\cite{Benini:2015noa, Benini:2016hjo,  Closset:2016arn, Closset:2017zgf}, and the equality of twisted indices for the other SQCD theories with $l=0$ then follows from standard RG flow arguments. Here, our focus was instead  on computing the index on both sides explicitly, and the fact that \eqref{Zg dual eq} indeed holds in many examples%
 \footnote{On a laptop computer, we can check the matching of indices accross dualities for most gauge theories with rank up to 3 and with the parameters $k, l, n_f, n_a$ small enough.} is a nice check of our formalism. Moreover, for unitary SQCD with $l\neq 0$, which we study in the next section, no general proof of this equality is available so far.

 \subsection{Special cases: abelian dualities}\label{subsec:abDuals}
Let us briefly discuss a few special cases with $N_c=1$ where the dual theory consists of chiral multiplets only. These `elementary' dualities will be particularly useful in the next section.

\medskip
\noindent
{\bf The SQED/$XYZ$ duality.} Consider the $U(1)_0$ theory with $n_f=n_a=1$. It has a dual description in terms of three chiral multiplets $M$, $\T^\pm$ with the superpotential:
\be
W= \T^+ \T^- M~,
\ee
also known as the $XYZ$ model (with $X=M$, $Y=\T^+$, $Z=\T^-$).
This can be viewed as a limiting case of Aharony duality. This theory has a flavour symmetry $U(1)_T\times U(1)_A$ and a $R$-symmetry. In our conventions, we have the following flavour CS contact terms in the dual description:
\bea	 \label{Kshifts SQED XYZ}
& K_{TT}  =  1~,\qquad && K_{TA}= K_{TR}=0~,\\
&	K_{AA} = 2~, \qquad && K_{AR}=2r~,\\
&	K_{RR} =2 r^2~, \qquad && K_{g}= 2~.
	\eea
as a limiting case of \eqref{Aharony-cs-levels}.

\medskip
\noindent
{\bf The $U(1)_{\pm 1}$ CS theory.} Consider the 3d $\CN=2$ CS theory $U(1)_1$ without matter fields. This is the `almost trivial theory' studied in \cite{Witten:2003ya}:  it is dual to an invertible theory ({\it i.e.} a trivial theory with CS contact terms):
\be\label{almost triv dual 1}
U(1)_1 \qquad \longleftrightarrow \qquad K_{TT}=-1~, \;\; K_{RT}=0~,\; \; K_{RR}=0~,\;\; K_{g}= -2~.
\ee
This is a special case of the Giveon-Kutasov duality~\cite{Giveon:2008zn}, and these CS contact terms are obtained by setting $N_c=n_a=n_f=k=1$ in table~\ref{tab: bare levels min chiral}. For the other sign of the Chern-Simons level, we similarly find:
\be\label{almost triv dual 2}
U(1)_{-1} \qquad \longleftrightarrow \qquad K_{TT}=1~, \;\; K_{RT}=0~,\; \; K_{RR}=1~,\; \; K_{g}= 4~.
\ee

\medskip
\noindent
{\bf The $U(1)_{\pm \half}$ theory coupled to one chiral flavour.} Let us consider the $U(1)_{\half}$ theory coupled to one chiral multiplet $\Phi_\pm$ of electric charge $\pm 1$ and $R$-charge $r$, with $k=\half$. This theory has a flavour symmetry $U(1)_T$, and it is dual to a free chiral multiplet $\T^\pm$ of $U(1)_T$ charge $\pm 1$ and $R$-charge $1-r$, with the following CS contact terms:
\be\label{elementary U12 dual 1}
U(1)_\half~, \; \Phi_\pm~, \; K_{GR}=0~, \quad  \longleftrightarrow \quad \T^\pm~, \; \; \begin{cases} K_{TT}=0~, \; \; &K_{RT}=\mp r~, \\ K_{RR}=r^2~, \; \; & K_g= 0~.\end{cases}
\ee
Note that, in our conventions, we have a bare CS level $K_{GG}=1$ for the $U(1)$ gauge group on the `electric' side of the duality.
With the opposite sign for the UV CS level, $k= -\half$, we find instead:
\be\label{elementary U12 dual 2}
U(1)_{-\half}~, \; \Phi_\pm~, \; K_{GR}=\pm(r-1)~, \quad  \longleftrightarrow \quad \T^\mp~, \; \; \begin{cases} K_{TT}=1~, \; \; &K_{RT}=0~, \\ K_{RR}=-r^2+2 r~, \; \; & K_g= 2~.\end{cases}
\ee
These well-known dualities~\cite{Dorey:1999rb} are limiting cases of the marginally chiral dualities reviewed in section~\ref{subsec:l0margchir}.

\section{Infrared dualities for $U(N_c)_{k, k+l N_c}$ SQCD}\label{sec:SQCDlnon0}
In this section, we discuss the infrared dualities for unitary SQCD with general value of $l$. These dualities were first discovered by Nii for $n_a=n_f$ and $k\neq 0$ \cite{Nii:2020ikd}. They were further generalised by Amariti and Rota \cite{Amariti:2021snj}, who argued that the dualities with $l\neq 0$ can be easily derived from the $l=0$ dualities by using Kapustin-Strassler and Witten's ${\rm SL}(2,\Z)$ action on 3d field theories with abelian symmetries~\cite{Kapustin:1999ha, Witten:2003ya}. We elaborate on this construction in the following.

\subsection{${\rm SL}(2,\Z)$ action and the 3d $A$-model}
Let us first discuss the Kapustin-Strassler-Witten ${\rm SL}(2,\Z)$ transformations in the language of the 3d $A$-model. We will then rederive all the $l\neq 0$ dualities by an appropriate ${\rm SL}(2,\Z)$ transformation of the $l=0$ dualities summarised in the previous section. We again pay particular attention to deriving the exact bare CS levels in all cases. 

 Let us  consider some 3d $\CN=2$ supersymmetric gauge theory $\CT$ with a $U(1)_f$ flavour symmetry,  and the associated 3d $A$-model determined by the twisted superpotential $\CW(u, \nu)$ and the effective dilaton $\Omega(u, \nu)$, with $\nu$ the $U(1)_f$ chemical potential ({\it i.e.} the 2d twisted mass). Here $u_a$ denotes gauge parameters for dynamical vector multiplets, and the remaining flavour parameters are left implicit.
The ${\rm SL}(2,\Z)$ action sends $\CT$ to another field theory $g[\CT]$ for any $g \in {\rm SL}(2,\Z)$. Let $\SW$ and $\TW$ denote the two standard generators of ${\rm SL}(2,\Z)$, with:
\be\label{SL2Z rels}
\SW^2 = \textbf{C}~, \qquad\qquad (\SW \TW)^3=\textbf{C}~, \qquad\qquad \textbf{C}^2={\bf 1}~,
\ee
where $\textbf{C}$ is the central element generating $\Z_2\subset {\rm SL}(2,\Z)$. These actions generate new field theories with the same number of $U(1)$ symmetries. They act on $\CT$ as follows:
\begin{itemize}

\item[(i)] $\SW \, : \, \CT \rightarrow \SW[\CT]\equiv \CT/U(1)_f$ corresponds to gauging the abelian symmetry $U(1)_f$ with a 3d $\CN=2$ vector multiplet. The new $U(1)_f$ dynamical field strength $F_f= dA_f$ gives us the conserved current of a new topological symmetry, denoted by $U(1)_{f'}$, which we couple to a background $U(1)_{f'}$ multiplet with a supersymmetric  $f$-$f'$  mixed CS level, also known as a 3d BF term:
\be\label{BF 3d S transfo}
- {i\ov 2 \pi} \int  \left(A_{f'} \wedge F_f + \cdots\right)~,
\ee
where the ellipsis denotes the supersymmetric completion. At the level of the 3d $A$-model, we rename $\nu$ as $v$ to indicate that it is now a gauge parameter, and the new coupling \eqref{BF 3d S transfo} appears as a quadratic term in the new twisted superpotential:
\be
\CW(u, \nu)  \overset{\SW}{\longrightarrow} \CW(u, v) - \nu' v~, \qquad\qquad \Omega(u, \nu)  \overset{\SW}{\longrightarrow}   \Omega(u, v)~.
\ee
with $\nu'$ the $U(1)_{f'}$ parameter. We then simply add a new equation for the $U(1)_f$ gauge symmetry to the Bethe equations:
\be
\left\{ \Pi_a\equiv e^{2\pi i {\d \CW \ov \d u_a}}=1\right\} \qquad \overset{\SW}{\longrightarrow}  \qquad \left\{ \Pi_a = 1~, \;  \Pi_{v}\equiv e^{2\pi i \left({\d \CW\ov \d v}- \nu' \right)}=1\right\}~.
\ee

\item[(ii)] $\TW \, : \, \CT \rightarrow \TW[\CT]$ corresponds to shifting the $U(1)_f$ CS contact term by $1$:
\be
\kappa_{ff} \rightarrow \kappa_{ff} +1~,
\ee
by adding a level-$1$ 3d $\CN=2$ supersymmetric Chern-Simons interaction for the $U(1)_f$ background vector multiplet to the action:
\be
S \rightarrow S+ {i\ov 4\pi}\int \left(A_f \wedge d A_f + \cdots \right)~.
\ee
In the 3d $A$-model, we then have:
\be
\CW(u, \nu)  \overset{\TW}{\longrightarrow} \CW(u, \nu)+ \half(\nu^2 + \nu) ~, \qquad\qquad \Omega(u, \nu)  \overset{\TW}{\longrightarrow}   \Omega(u, \nu)~.
\ee

\item[(iii)]  The central element $\textbf{C}$ acts as sign flip on the $U(1)_f$ current and its superpartners, which is equivalent to a sign flip of the $U(1)_f$ background vector multiplet. Thus, in the 3d $A$-model:
\be
\CW(u, \nu)  \overset{\textbf{C}}{\longrightarrow} \CW(u, -\nu)~, \qquad\qquad \Omega(u, \nu)  \overset{\textbf{C}}{\longrightarrow}   \Omega(u, -\nu)~.
\ee
\end{itemize}

\noindent
It is interesting to verify the ${\rm SL}(2,\Z)$ relations \eqref{SL2Z rels} directly in the 3d $A$-model formalism. We use the fact that, when a gauge field $A_0$   only appears linearly through a 3d BF term, 
\be
S_0 = {i\ov 2\pi} \sum_{i\neq 0} K_{0 i} \int A_0 \wedge d A_i~,
\ee
the path integral over $A_0$ gives us a functional Dirac $\delta$-function \cite{Witten:2003ya}:
\be
\int [dA_0] e^{-S_0} = \delta\big( \sum_{i\neq 0} K_{0 i} A_i\big)~,
\ee
and similarly in the 3d $\CN=2$ supersymmetric context. 
 Now, consider the action:
\be
\SW^2\; :\; \CW(\nu)\longrightarrow \CW( \nu'')= \CW(v)- v' v -\nu'' v'~.
\ee
Here, the path integral over the  $v'$ vector multiplet gives us $\delta(v+ \nu'')$, schematically speaking, and therefore we obtain the original theory with a sign flip of $\nu$, as expected:
\be
\CW(u, \nu)  \overset{\SW^2=\textbf{C}}{\longrightarrow} \CW(u, -\nu'')~,
\ee
up to a slight subtlety to be discussed momentarily.
To compute  $(\SW \TW)^3$, note that we have:
\be
\SW \TW \; :\; \CW(u,\nu)\longrightarrow   \CW(u,v)+ \half v(v+1)-\nu' v~,
\ee
and therefore:
\be
 \CW(u,\nu)\overset{(\SW \TW)^3}{\longrightarrow}  \CW(u,v)+ \half v(v+1)  + {\half v' (v'+1)+ \half v''(v''+1) -v' v - v'' v' - \nu''' v''}~.
\ee
After performing a change of variable $v'\rightarrow v'+v''+\nu'''$, integrating out $v''$ gives us  $\delta(v+ \nu''')$, and we obtain:
\be\label{ST3 action}
 \CW(u,\nu)\overset{(\SW \TW)^3}{\longrightarrow}  \CW^{\CT}(u, -\nu''')+ \half v'(v'+1)+ 2 \nu''' (\nu'''+v')~.
\ee
The (sign-flipped) original theory is now tensored with a decoupled topological sector which is an `almost trivial' theory~\cite{Witten:2003ya}, namely a $U(1)_1$ CS theory. Indeed, the additional Bethe equation for $v'$ in \eqref{ST3 action} is decoupled from the other Bethe equations  of the full theory, and it has a unique solution.

\subsection{From $U(N_c)_k$ to $U(N_c)_{k, k+l N_c}$ SQCD}

The ${\rm SL}(2,\Z)$ action allows us to generate a non-zero CS level $l$ starting from a $U(N_c)_k$ gauge theory, as we now explain. 

\medskip
\noindent
{\bf $\SW$ and $\SW^{-1}$ on 3d $\CN=2$ supersymmetric theories.} When acting with $\SW$ on $\CT$, we introduce a new abelian vector multiplet. It contains a single gaugino, which shifts some of CS contact terms in the UV according to:
\be
\kappa_{RR}\overset{\SW}{\rightarrow} \kappa_{RR}+\half~, \qquad \quad \kappa_{g}\overset{\SW}{\rightarrow} \kappa_{g}+1~,
\ee
in our conventions. Thus, more precisely, the action of $\SW^2$ on $\CT$ actually gives us:
\be
\CW(u, \nu)  \overset{\SW^2=\textbf{C} }{\longrightarrow} \CW(u, -\nu'') +{1\ov 12}~, \qquad 
\Omega(u, \nu)  \overset{\SW^2=\textbf{C} }{\longrightarrow} \Omega(u, -\nu'') +{1\ov 2}~,
\ee
which includes the shifts $K_g\rightarrow K_g+2$ and $K_{RR}\rightarrow K_{RR}+1$. By a slight abuse of notation,  let us then define an inverse operation:
\be\label{defSinv i}
\SW^{-1} \equiv \delta(K) \circ  \textbf{C} \circ \SW~.
\ee
It consists of the naive inverse, $ \textbf{C} \circ \SW$, combined with a shift of the bare CS levels:
\be\label{defSinv ii}
 \delta(K) \; : \; K_{RR}\rightarrow K_{RR}-1~, \qquad K_g \rightarrow K_g-2~,
\ee
so that $\SW^{-1} \SW$ is truly the identity on $\CT$.

\medskip
\noindent
{\bf The $\SW^{-1}\TW^l \SW$ action on $U(N_c)_k$ SQCD.} Let us now start with $\CT$ being SQCD with $l=0$. We can obtain the $l\neq 0$ theory by acting with $\SW^{-1}\TW^l \SW$ on the topological symmetry $U(1)_T$ of the $l=0$ theory. Indeed, at the level of the $A$-model, let $\tau$ denote the $U(1)_T$ parameter and $\nu$ the other flavour parameters. Let us also decompose the gauge parameters $u_a$ as 
\be\label{def ua u0}
u_a= \t u_a + u_0~, \qquad \quad \sum_{a=1}^{N_c} \t u_a=0~.
\ee
  We have that the twisted superpotential is linear in $\tau$:
\be
\CW(u, \nu, \tau) = \CW_0(u, \nu) + \tau N_c u_0~, 
\ee
and that $\Omega= \Omega(u,\nu)$ is $\tau$-independent. That is, all the flavour CS levels $K_{T\alpha}$ and $K_{RT}$ vanish, with $\tau$ coupling to the gauge symmetry with $K_{GT}=1$. To act with $\SW^{-1}\TW^l \SW$, we first render $U(1)_T$ dynamical, relabelling $\tau \rightarrow v$, and we introduce $w$ the flavour parameter for the new topological symmetry. We add a level-$l$ for the latter, before gauging it with $\SW^{-1}$, and we call the new abelian flavour symmetry $U(1)_T$ again, with a new parameter $\tau$. Thus, we have:
\be
\CW(u, \nu, \tau)\overset{\SW^{-1}\TW^l \SW}{\longrightarrow} \CW_0(u, \nu) + v N_c u_0 - w v+ {l\ov 2} w(w+1) + \tau w~. 
\ee
The vector multiplet for $v$ only appears linearly, hence we can integrate it out, which leads to a $\delta$-function constraint $w= N_c u$, and we then obtain precisely the general SQCD theory:
\be
\CW(u, \nu, \tau)\overset{\SW^{-1}\TW^l \SW}{\longrightarrow}  \CW(u, \nu, \tau)+ {l\ov 2} N_c u_0(N_c u_0+1)~.
\ee
This action only introduced the $l\neq 0$ CS term, and it did not change any of the flavour CS levels thanks to the definition~\eqref{defSinv i}-\eqref{defSinv ii}.

\subsection{Amariti-Rota duality ($k=0$, $n_f=n_a$)}\label{subsec:AR dual}
To obtain the dual descriptions of SQCD with generic $l$, we can simply act with  $\SW^{-1}\TW^l \SW$ on the dual descriptions reviewed in the previous section. Let us start with the case of $k=0$ and $n_f=n_a\equiv N_f$. The dual description at $l=0$ is the Aharony magnetic theory discussed below \eqref{aharony dual}. 
At the level of the 3d $A$-model, the $\SW^{-1}\TW^l \SW$ action on the Aharony dual theory gives us:
\bea
&\CW&=&\;\; \CW_0+ {1\ov (2\pi i)^2}\left(\dilog(z y_A^{-N_f}) + \dilog(z^{-1}y_A^{-N_f}) \right) - v N_c^D u_0 + \half v(v+1) - v w\\
&&&\;\; + {l\ov 2} w(w+1) + \tau w~,
\eea
where we renamed $\tau$ to $v$ (and $q=e^{2\pi i \tau}$ to $z= e^{2\pi i v}$),  $\CW_0$ is $v$-independent, and we used the same notation as in \eqref{def ua u0} for the dual gauge group $U(N_c^D)$.  This new theory contains a subsector that is isomorphic to SQED. Indeed, we have:
\bea\label{CW decompose WSQED}
&\CW= \CW_0  + {l\ov 2} w(w+1) + \tau w + \CW_{\rm SQED}~,\\
& \CW_{\rm SQED}\equiv {1\ov (2\pi i)^2}\left(\dilog(z y_A^{-N_f}) + \dilog(z^{-1}y_A^{-N_f}) \right)  + \half v(v+1) + \t \tau v~,
\eea
with $\t\tau \equiv - w - N_c^D  u_0$. Using the SQCD/$XYZ$ duality reviewed in section~\ref{subsec:abDuals}, we can integrate out the vector multiplet for $v$, and we obtain:
\bea\label{WSQED dual}
& \CW_{\rm SQED}& \leftrightarrow&\;\;\; {1\ov (2\pi i)^2}\left(\dilog( y_A^{-2N_f}) +  \dilog(y_A^{N_f} x_0 x_{(w)} ) + \dilog(y_A^{N_f} x_0^{-1} x_{(w)}^{-1} ) \right)\\
  &&&\; \;+ N_f^2 \nu_A(\nu_A+1) + \half w(w+1) + \half N_c^D u_0(N_c^D u_0+1) + N_c^D w u_0+ {1\ov 12}~,
\eea  
and similarly for the effective dilaton. Here $x_0= e^{2\pi i u_0}$ and $x_{(w)}= e^{2\pi i w}$. The local application of the duality shifts various flavour CS terms, as dictated by \eqref{Kshifts SQED XYZ}, and one must also take into account the shift \eqref{defSinv ii}. In total, only the  $K_{AR}$ and $K_{RR}$ CS contact terms incur a shift $K \rightarrow K+ \Delta K$ with respect to the Aharony dual theory, with:
\be\label{Ah to AR shift}
\Delta K_{AR}=- N_f(2+ 2 (r_T-1))~, \qquad \Delta K_{RR} = 4(r_T-1) +2 (r_T-1)^2 +1~.
\ee
Moreover, the new topological symmetry $U(1)_T$ corresponds to the magnetic flux of the new $U(1)^{(w)}$ gauge group (with vector multiplet $w$), and the new FI term $\tau$ enters as in \eqref{CW decompose WSQED}, thus we have $K_{Tw}=1$. Proceeding in this way, we obtain the Amariti-Rota duality \cite{Amariti:2021snj}:
\be\label{AR dual}\boxed{
U(N_c)_{0, l N_c}~, \; N_f  ({\tiny\yng(1)}\oplus \overline{ {\tiny\yng(1)}})  \qquad
\longleftrightarrow \qquad
\begin{matrix} U(N_f- N_c\underbracket[0.7pt][7pt]{)_{0, \,0} \times U(1}_0)^{(w)}_{l}~,\\
\\
 \; N_f  ({\tiny\yng(1)}\oplus \overline{ {\tiny\yng(1)}})~,  \; \left({M_i}^j, \,\CB_+,\,  \CB_-, \, X\right).
 \end{matrix}}
\ee
The dual matter fields and their charges are given in table~\ref{tab: AR duality charges}.
\begin{table}[t]
\renewcommand{\arraystretch}{1.1}
\centering
\be\nn
\begin{array}{|c|cc|ccccc|}
\hline
    & U(N_c^D)&U(1)^{(w)}& SU(N_f) & SU(N_f)  & U(1)_A &  U(1)_T & U(1)_R  \\
\hline
q_j   &{\tiny\yng(1)}&0&    \bm{1} & \overline{\tiny\yng(1)} & -1   & 0   &1-r \\
\t q^i       &\overline{\tiny\yng(1)}&0 &{\tiny\yng(1)}   & \bm{1}    & -1   & 0   &1-r \\
{M ^j}_i      & \bm{1} &0&\overline{\tiny\yng(1)}& {\tiny\yng(1)} &  2   & 0   &2 r \\
\CB_+       &  {\rm det}^{+1} &1& \bm{1} & \bm{1} & N_f   & 0   & -r_T+1\\
\CB_-     & {\rm det}^{-1}  &-1& \bm{1} & \bm{1} & N_f   & 0   & -r_T+1 \\
X     & \bm{1} &0&\bm{1}&\bm{1}&  -2 N_f   & 0   &2 r_T \\
\hline
\end{array}
\ee
\caption{Field content of the Amariti-Rota dual theory, with $N_c^D= N_f-N_c$. Here ${\rm det}^{\pm 1}$ denotes the one-dimensional representation of $U(N_c^D)$ with weight $\rho=(\pm 1, \cdots, \pm 1)$,  and $r_T= -N_f(r-1)-N_c+1$ like in the Aharony dual theory.}
\label{tab: AR duality charges}
\end{table}
 In addition to the dual flavours and mesonic fields, we have an additional singlet $X$ as well as the `baryonic' fields $\CB_\pm$ charged under both $U(1)$ factors of the gauge group. The gauge singlets are coupled to the gauge sector through the superpotential:
\be
W=\t q^i {M_i}^j q_j + \CB_+ \CB_- X~.
\ee
Note that the $U(1)^{(w)}$ gauge group has an UV effective CS level $l$, which corresponds to $K_{ww}= l+1$. Similarly, the effective mixed CS level between the two gauge groups vanishes, $k_{G_0 w}=0$, but we have a bare CS level $K_{G_0 w}=1$ as shown in \eqref{WSQED dual}.
Finally, the flavour bare CS levels for the Amariti-Rota dual are slightly different from the ones of the Aharony dual theory due to \eqref{Ah to AR shift}. We have:
\bea\label{ar-cs-levels}
	&K^{(\text{AR})}_{SU(N_f)} = \tilde{K}^{(\text{AR})}_{SU(N_f)} = N_{c}^D~,\\
	&K^{(\text{AR})}_{AA} = 2N_f \left(N_c^D + 2N_f\right)~,\\
	&K^{(\text{AR})}_{AR} = -2N_f\left(2N_c^D+1\right) + 2N_f\left(N_c^D+2N_f\right)r~,\\
	& K^{(\text{AR})}_{RR} = 1 + N_c^D(3 N_c^D+4) -4N_f\left(2N_c^D+1\right)r + 2N_f\left(N_c^D + 2N_f\right)r^2~,\\
	&K_{g}^{(\text{AR})} = 2N_f N_c^D+2~,
\eea
with all other flavour CS levels vanishing.

\subsection{Minimally chiral duality with general $l$}
Let us now turn on the CS level $k$. We use the notation $k_c \equiv \half (n_f- n_a)$ as before, and we first consider the minimally chiral case, namely the case $|k| > |k_c|$ with $k\neq 0$.
 The dual theory is obtained by an $\SW^{-1}\TW^l \SW$ action on the dual theory of section~\ref{subsec:l0minchir}. In the 3d $A$-model, this gives us:
 \be
 \CW= \CW_0 - v N_c^D u_0 - {\sign(k)\ov 2} v (v+1) - v w + {l\ov 2}w(w+1) + \tau w~,
 \ee
with the same conventions as in the previous subsection. Now, the subsector involving the gauge field for $v$ is simply the `almost trivial' CS theory $U(1)_{-\sign(k)}$. We use the duality \eqref{almost triv dual 1} if $k<0$, and  we use the duality \eqref{almost triv dual 2} if $k>0$. 
By a straightforward computation, we then derive the following generalised Nii duality~\cite{Nii:2020ikd, Amariti:2021snj}: 
\be\label{min chiral dual genl}
\boxed{
U(N_c)_{k, k+ l N_c}~, \;  \left(n_f\, {\tiny\yng(1)}, n_a\, \overline{ {\tiny\yng(1)}}\right)  \quad
\longleftrightarrow \quad
\begin{matrix} U(N_c^D\underbracket[0.7pt][7pt]{)_{-k, \, -k +\sign(k) N^D_c} \times U(1}_{\sign(k)})_{l+\sign(k)}~,\\
\\
 \; \left(n_a\, {\tiny\yng(1)}, n_f\, \overline{ {\tiny\yng(1)}}\right)~,  \; \left({M_i}^j\right).
 \end{matrix}
 }
\ee
with $N_c^D= |k|+ \half(n_f+n_a)-N_c$ and the dual superpotential $W=\t q_i {M_i}^j q_j$. 

The matter content of the magnetic theory is given in the upper part of table~\ref{tab: general dual genl}. 
\begin{table}[t]
\renewcommand{\arraystretch}{1.1}
\centering
\be\nn
\begin{array}{|c|cc|ccccc|c|}
\hline
    & U(N_c^D)&U(1)^{(w)}& SU(n_f) & SU(n_a)  & U(1)_A &  U(1)_T & U(1)_R & \text{condition} \\
\hline
q_j   &{\tiny\yng(1)}&  0&  \bm{1} & \overline{\tiny\yng(1)} & -1   & 0   &1-r &\\
\t q^i       &\overline{\tiny\yng(1)}&0&{\tiny\yng(1)}   & \bm{1}    & -1   & 0   &1-r &\\
{M ^j}_i      & \bm{1} &0&\overline{\tiny\yng(1)}& {\tiny\yng(1)} &  2   & 0   &2 r &\\
\hline
\CB_+       &  {\rm det}^{+1} &1& \bm{1} & \bm{1} & N_f   & 0   & -r_T+1 &   |k|= k_c \\
\hline
\CB_-     & {\rm det}^{-1}  &-1& \bm{1} & \bm{1} & N_f   & 0   & -r_T+1  &   \;  |k|= -k_c  \\
\hline
\end{array}
\ee
\caption{Field content of the infrared dual of unitary SQCD with $l=0$ and $|k|\geq |k_c|$. The baryonic fields $\CB_\pm$ only appear in the marginally chiral case (or in the Amariti-Rota dual), as indicated. Recall that $r_T= -(N_c+N_c^D)(r-1)-N_c+1$.}
\label{tab: general dual genl}
\end{table}
As for the Amariti-Rota duality, the new $U(1)_T$  symmetry only enters through the standard FI term of the $U(1)^{(w)}$ gauge group, and the other mixed gauge-global CS levels (not involving $U(1)_T$)  remain the same as in \eqref{mixed GF min chiral}, hence we have:
\be \label{mixed GT terms gen l min chir}
K_{G_0T}=0~,\quad K_{wT}=1~, \quad K_{GA} = \Theta(k) (n_f-n_a)~, \quad K_{GR} = \Theta(k) (n_f-n_a)(r-1)~.
\ee
The flavour bare CS levels of the generalised Nii dual are given in table~\ref{tab: bare levels min chiral genl}. 

Note that for $l=0$ the second gauge group $U(1)_{l\pm 1}$ reduces to $U(1)_{\pm 1}$. This can be eliminated from the description using a local duality,  thus recovering the dual theory of section~\ref{subsec:l0minchir}. 
\begin{table}
\renewcommand{\arraystretch}{1.2}
\centering
\be\nn
\begin{array}{|c|c|c|}
\hline
    & k > |k_c| & k <  - |k_c|\\
\hline\hline
\; K_{SU(n_f)} \;&N_c^D& - N_c + n_a\\
\hline
\; K_{SU(n_a)} \;&N_c^D& - N_c + n_f\\
\hline
\; K_{AA} \;& (n_f+n_a) N_c^D&  4 n_f n_a -(n_f+n_a) N_c\\
\hline
\; K_{TT} \;& 0 & 0 \\
\hline
\; K_{AT} \; & 0 & 0\\
\hline
\; K_{RA}\;& (n_f+n_a) N_c^D r &  (n_a+n_f)N_c - 2 n_f n_a+ (4 n_f n_a - (n_f+n_a) N_c)r\\
\hline
\; K_{RT} \;& 0 & 0\\
\hline
\multirow{3}{*}{\; $K_{RR}$ \;}&\multirow{3}{*}{\quad $(-N_c^D + (n_f+n_a)r^2)N_c^D$\quad} &  (N_c-n_f)(N_c-n_a)-1\\
   & & +2 r ((n_a+n_f)N_c - 2 n_f n_a)   \\
   & &  +r^2 (4 n_f n_a -(n_f+n_a) N_c)  \\
\hline
\; K_{g} \;&  (n_f + n_a - 2k)N_c^D &  2 n_f n_a -(n_f+n_a+2 k) N_c -2\\
\hline
\end{array}
\ee
\caption{Flavour bare CS levels for the minimally-chiral dual theory with general $l$.}
\label{tab: bare levels min chiral genl}
\end{table}
 
\subsection{Marginally chiral duality with general $l$}
Next, we consider the marginally chiral case, $|k|= |k_c|$. It is most convenient to consider the cases with positive and negative $k$ separately.

\medskip
\noindent
{\bf Marginally chiral case with $k=|k_c|$.} For definiteness, let us start by considering the case $k=k_c>0$ (with $n_f >n_a$). The  $\SW^{-1}\TW^l \SW$ transformation on the magnetic theory of section~\ref{subsec:l0margchir} gives us:
\be
\CW= \CW_0 + {1\ov (2\pi i)^2} \dilog(y_A^{-n_f} z)- v N_c^D u_0 - n_f v \nu_A - w v  + {l\ov 2} w(w+1) + \tau w~,
\ee
where again $\CW_0$ denotes all the $\tau$-independent terms in twisted superpotential the theory we started with. Renaming $\tau \rightarrow v$ upon gauging $U(1)_T$,  the $v$-dependent terms precisely corresponds to the electric theory in the elementary duality \eqref{elementary U12 dual 2}, namely:
\be
\CW_{U(1)_{-\half}, \Phi_+} = {1\ov (2\pi i)^2} \dilog(y_A^{-n_f} z)+ \t \tau v~, \qquad \t \tau \equiv-  N_c^D u_0-w - n_f v \nu_A~.
\ee
More precisely, we can identify the singlet $\T^+$ with the field $\Phi_+$ in \eqref{elementary U12 dual 2}, with $R$-charge $r\rightarrow r_T$, noting that $K_{RT}= r_T-1$ in original $l=0$ theory. 
The path integral over the vector multiplet $v$ then gives a dual singlet which we can call $\CB_+$, of $R$-charge $1-r_T$, with charge $({\rm det}, 1)$ under the remaining gauge group $U(N_c^D)\times U(1)^{(w)}$. We have:
\be
\CW_{U(1)_{-\half}, \Phi_+} \;\;\; \leftrightarrow\;\; {1\ov (2\pi i)^2} \dilog(x_0^{N_c^D} x_{(w)} y_A^{n_f}) + \half \t\tau (\t\tau +1)+{1\ov 12}~.
\ee
  The twisted superpotential of the new dual theory then reads:
\bea
&\CW&=&\;\; \CW_0  + {1\ov (2\pi i)^2} \dilog(x_0^{N_c^D} x_{(w)} y_A^{n_f}) + \half N_c^D u_0(N_c^D u_0+1)  + N_c^D u_0 w\\
&&&\;\qquad + {l+1\ov 2} w(w+1)+ \tau w - {n_f^2\ov 2} \nu_A(\nu_A+1)~,
\eea
and we have a similar transformation of the effective dilaton,  as dictated by the dual flavour CS levels of the elementary duality \eqref{elementary U12 dual 2}. 
A careful accounting of the bare CS levels gives us the following non-zero shifts $K \rightarrow K+\Delta K$ with respect to the $l=0$ theory:
\be\label{shift marg chiral case 1}
\Delta K_{AA}= -n_f^2~,\quad \Delta K_{RA}= (r_T-1)n_f~, \quad \Delta K_{RR}= - r_T^2+ 2 r_T -1~.
\ee 
A similar computation can be carried out for the case $k= - k_c >0$ (with $n_a >n_f$). This replaces the singlet $\T_-$ in the original theory with a chiral multiplet $\CB_+$ of charge $({\rm det}^{-1}, -1)$ under $U(N_c^D)\times U(1)^{(w)}$. The flavour CS level shifts are like in \eqref{shift marg chiral case 1} with $n_f$ replaced by $n_a$.

\begin{table}
\renewcommand{\arraystretch}{1.2}
\centering
\be\nn
\begin{array}{|c|c|c|}
\hline
    & k = |k_c|>0  & k = - |k_c|<0\\
\hline\hline
\; K_{SU(n_f)} \;&N_c^D&- N_c + n_a  \\
\hline
\; K_{SU(n_a)} \;&N_c^D&- N_c + n_f  \\
\hline
\; K_{AA} \;& (n_f+n_a)N_c^D & (n_f+ n_a) N_c^D + 3 n_a n_f \equiv K_{AA}^{(0)}  \\
\hline
\; K_{TT} \;&0 & 0\\
\hline
\; K_{AT} \; &0& 0\\
\hline
\multirow{3}{*}{\; $K_{RA}$ \;} &\;\; \multirow{3}{*}{\; $N_c^D (n_f +n_a)r$}  \;\; &\;\; K_{RA}^{(0)}+r K_{AA}^{(0)}~,    \;\; \\
&&  K_{RA}^{(0)}\equiv  -N_c^D (n_f+n_a) + 2\max(n_f, n_a)^2  \;\; \\
&&\qquad\qquad-\max(n_f, n_a)(N_c^D + n_f +n_a +1) \quad  \\
\hline
\; K_{RT} \;& 0  \;\; &\;\;  0  \;\; \\
\hline
\multirow{2}{*}{\; $K_{RR}$ \;}& \multirow{2}{*}{\; $N_c^D (-N_c^D + (n_f+n_a)r^2)$ }&  N_c^D(2 N_c^D + n_f+n_a-2 \max(n_f, n_a)+2)    \\
&& 2 r K_{RA}^{(0)}+ r^2 K_{AA}^{(0)} \\
\hline
\; K_{g} \;& 2 \max(n_f, n_a) N_c^D & 2 N_c^D (n_f+n_a- \max(n_f, n_a))  \\
\hline
\end{array}
\ee
\caption{Flavour bare CS levels for the marginally chiral dual theories,  $|k|= |k_c|$, for general $l$.}
\label{tab: bare levels marg chiral genl}
\end{table}
\medskip
\noindent
In summary, for $k= \epsilon k_c >0$, for $\epsilon= \pm$, we have the duality:
\be\label{gen marg dual 1} \boxed{
U(N_c)_{k, k+ l N_c}~, \;  \left(n_f\, {\tiny\yng(1)}, n_a\, \overline{ {\tiny\yng(1)}}\right) \qquad
\longleftrightarrow \qquad
\begin{matrix} U(N_c^D\underbracket[0.7pt][7pt]{)_{-k, \,-k +\half N_c^D} \times U(1}_{\half})^{(w)}_{l+\half}~,\\
\\
 \; \left(n_a\, {\tiny\yng(1)}, n_f\, \overline{ {\tiny\yng(1)}}\right)~,  \; \left({M_i}^j, \,\CB_\epsilon \right)~,
 \end{matrix}
}
\ee
with the dual superpotential $W=\t q_i {M_i}^j q_j$. The precise matter content is given in table~\ref{tab: general dual genl}. The mixed gauge-flavour CS levels are the same as in \eqref{mixed GT terms gen l min chir}, namely:
\be
K_{G_0T}=0~,\quad K_{wT}=1~, \quad K_{GA} =  (n_f-n_a)~, \quad K_{GR} =  (n_f-n_a)(r-1)~.
\ee
 and the bare flavour CS levels are given on the left-hand-side of table~\ref{tab: bare levels marg chiral genl}.

\medskip
\medskip
\noindent
{\bf Marginally chiral case with $k=-|k_c|$.} The dual theories with $k= -|k_c|<0$ can be derived similarly. In this case, we need to use the elementary duality \eqref{elementary U12 dual 1}. We find non-trivial shifts of the gauge-flavour CS levels:
\be
\Delta K_{G_0 A}=\Delta K_{w A} = \pm \max(n_f, n_a)~, \qquad
\Delta K_{G_0 R}=\Delta K_{w A} = \mp r_T~,
\ee
for $k= \mp k_c <0$. We also have the flavour CS levels shifts:
\bea
&\Delta K_{AA}=  \max(n_f, n_a)^2~, \qquad &&\Delta K_{AR} = - r_T \max(n_f, n_a)~,\\
& \Delta K_{RR}= r_T^2-1~, \quad && \Delta K_g = -2~,
\eea
compared to the levels shown on the right-hand-side of table~\ref{tab: bare levels min chiral}. 

\medskip
\noindent
Fixing $k= -\epsilon k_c <0$, for $\epsilon= \pm$, we then have the duality:
\be\label{gen marg dual 2}\boxed{
U(N_c)_{k, k+ l N_c}~, \;  \left(n_f\, {\tiny\yng(1)}, n_a\, \overline{ {\tiny\yng(1)}}\right) \qquad
\longleftrightarrow \qquad
\begin{matrix} U(N_c^D\underbracket[0.7pt][7pt]{)_{-k, \,-k -\half N_c^D} \times U(1}_{-\half})^{(w)}_{l-\half}~,\\
\\
 \; \left(n_a\, {\tiny\yng(1)}, n_f\, \overline{ {\tiny\yng(1)}}\right)~,  \; \left({M_i}^j, \,\CB_{\epsilon} \right)~,
 \end{matrix}
}
\ee
with the  superpotential $W=\t q_i {M_i}^j q_j$ and the matter content of table~\ref{tab: general dual genl}. We have the mixed gauge-flavour CS levels:
\be
K_{G_0 T}=0~, \quad  K_{wT}=1~, \quad K_{G_0 A}= K_{w A}= \epsilon \max(n_f, n_a)~, \quad 
K_{G_0 R}= K_{w A} =-\epsilon\,  r_T~,
\ee
with $r_T$ defined as in  table~\ref{tab: general dual genl}. The bare flavour CS level are given in table~\ref{tab: bare levels marg chiral genl}.

%

\subsection{Maximally chiral duality with general $l$}
Last but not least, let us consider the maximally chiral duality. Starting from the $l=0$ dual theory of section~\ref{subsec:l0maxgchir}, the $\SW^{-1}\TW^l \SW$ action gives us:
\bea
&\CW= \CW_0 -v N_c^D u_0 + K_{AT} ^{(\ast)} v \nu_A - v w + {l\ov 2} w(w+1) + \tau w~,\\
&\Omega= \Omega+ K ^{(\ast)}_{RT} v~.
\eea
Here, as before, $\CW_0$ and $\Omega_0$ contains all the $v$- and $w$-independent terms. Here we denote by $K^{(\ast)}$ the levels given in table~\ref{tab: bare levels max chiral}. Note that the vector multiplet for $v$ appears linearly, unlike in the previous cases. Therefore, integrating it out generate a functional $\delta$-function:
\be
\delta\left(W_\mu+ \tr(A_\mu) - K_{AT} ^{(\ast)} A_\mu^{(A)}-K_{AR} ^{(\ast)} A^{(R)}_\mu\right),
\ee 
with $\tr(A)= N_c^D A_0$ the $U(1)\subset U(N_c^D)$ gauge field, $W_\mu$ the $w$ gauge field, and $A_\mu^{(F)}$ denoting the background gauge fields for a $U(1)_F$ symmetry.  Note the appearance of the $U(1)_R$ gauge field because of the mixed topological-$R$ level, $K_{RT} ^{(\ast)}\neq 0$. Eliminating $w$ from the description, we obtain:
\bea
&\CW= \CW_0 - \tau N_c^D u_0 - \Delta K_{GA}\, \nu_A N_c^D u_0 + {l\ov 2} N_c^D u_0(N_c^D u_0+1) + \Delta K_{AA}\, \nu_A(\nu_A+1)~,\\
&\Omega=  \Omega_0 + \Delta K_{GR}\, N_c^D u_0+ \Delta K_{RA}\, \nu_A +\half \Delta K_{RR}~,
\eea
where we defined:
\bea
& \Delta K_{GA}= - l K_{AT} ^{(\ast)}~, \qquad && \Delta K_{GR}= - l K_{RT} ^{(\ast)}~, \\
& \Delta K_{AA}= l \left(K_{AT} ^{(\ast)}\right)^2~, \qquad && \Delta K_{RR}=  l \left(K_{RT} ^{(\ast)}\right)^2~, \\
& \Delta K_{RA}= l  K_{RT} ^{(\ast)}  K_{AT} ^{(\ast)}~.
\eea
In particular, we find that the maximally chiral dual theory has a gauge group $U(N_c^D)_{-k, -k +l N_c^D}$ with $N_c^D= \max(n_f, n_a)- N_c$, with the   matter content of table~\ref{tab: max chir dual gen 0}.
\begin{table}[t]
\renewcommand{\arraystretch}{1.1}
\centering
\be\nn
\begin{array}{|c|c|ccccc|}
\hline
    & U(N_c^D)& SU(n_f) & SU(n_a)  & U(1)_A &  U(1)_T & U(1)_R  \\
\hline
q_j   &{\tiny\yng(1)}&    \bm{1} & \overline{\tiny\yng(1)} & -1   & 0   &1-r \\
\t q^i       &\overline{\tiny\yng(1)}&{\tiny\yng(1)}   & \bm{1}    & -1   & 0   &1-r \\
{M ^j}_i      & \bm{1} &\overline{\tiny\yng(1)}& {\tiny\yng(1)} &  2   & 0   &2 r \\
\hline
\end{array}
\ee
\caption{Field content for the infrared dual of unitary SQCD with $|k|<|k_c|$.}
\label{tab: max chir dual gen 0}
\end{table}

\medskip
\noindent
In summary, we have:
\be\label{gen max chiral dual}
\boxed{
U(N_c)_{k,\,k+l N_c}~, \;  \left(n_f\, {\tiny\yng(1)}, n_a\, \overline{ {\tiny\yng(1)}}\right)  \quad
\longleftrightarrow \quad
U(N_c^D)_{-k, \, -k + l N_c^D}~, \;   \left(n_a\, {\tiny\yng(1)}, n_f \,\overline{ {\tiny\yng(1)}}\right)~,  \; \left({M_i}^j\right)~,}
\ee
with the usual Seiberg-dual superpotential. The dual theory contains the mixed gauge-flavour CS levels:
\bea\label{mixed GF max chiral gen l}
& K_{GT}=-1~,\\
&K_{GA} =\sign(k_c) \big(k+|k_c|+ l \max(n_f, n_a)\big)~, \\ 
&K_{GR} = \sign(k_c) \big(\big(k+|k_c|+l \max(n_f, n_a)  \big)(r-1)-  l N_c \big)~,
\eea
and the flavour bare CS levels given in table~\ref{tab: bare levels max chiral gen l}. 
\begin{table}
\renewcommand{\arraystretch}{1.2}
\centering
\be\nn
\begin{array}{|c|c|c|}
\hline
    & |k| < |k_c|~,\;\; n_f >n_a & |k| < |k_c|~,\;\; n_a >n_f \\
\hline\hline
\; K_{SU(n_f)} \;& k +\half(n_f+n_a)-N_c& n_a-N_c\\
\hline
\; K_{SU(n_a)} \;&n_f-N_c&  k +\half(n_f+n_a)-N_c\\
\hline
\multirow{2}{*}{\;$K_{AA}$\;}& K_{AA}^+\equiv (l+1)n_f^2+  3 n_f(k-k_c) &K_{AA}^- \equiv  (l+1)n_a^2+ 3 n_a(k+k_c)  \\
& + 2 N_c^D(n_f-k_c) &+ 2 N_c^D(n_a+k_c) \\
\hline
\; K_{TT} \;&0 & 0\\
\hline
\; K_{AT} \; &- n_f & n_a\\
\hline
\multirow{3}{*}{\;$K_{RA}$\;}&\;\; K_{RA}^{(0)+}+ r K_{AA}^+~,  \;\; &\;\;  K_{RA}^{(0)-}+ r K_{AA}^-~,\;\; \\
&K_{RA}^{(0)+} \equiv  -  (k-k_c)(n_f+ N_c^D) & K_{RA}^{(0)-} \equiv -  (k+k_c)(n_a+ N_c^D) \\
& \qquad\quad+ (l+1)n_f N_c^D& \qquad\quad+ (l+1)n_a N_c^D \\
\hline
\; K_{RT} \;& N_c^D- r n_f&-N_c^D+r n_a  \\
\hline
 \; K_{RR} \;& N_c^D (k-k_c+ l N_c^D)+ 2 r K_{RA}^{(0)+} + r^2 K_{AA}^+  &  N_c^D (k+k_c+ l N_c^D)   + 2 r K_{RA}^{(0)-} + r^2 K_{AA}^-  \\
\hline
\; K_{g} \;& 2 n_a  N_c^D & 2 n_f  N_c^D\\
\hline
\end{array}
\ee
\caption{Flavour bare CS levels for the maximally chiral dual, $|k|< |k_c|$, for general $l$. Note that, unlike in all the other cases, the flavour CS levels depend on $l$.}
\label{tab: bare levels max chiral gen l}
\end{table}
%
This dual theory trivially reduces to the dual of section~\ref{subsec:l0maxgchir} upon setting $l=0$. 

\medskip
\noindent
As a special case of the maximally chiral duality, consider $n_a=0$ and a CS level $|k|< {n_f\ov 2}$. We then have the simple-looking duality:
\be\boxed{
U(N_c)_{k,\,k+l N_c}~, \;   n_f \,  {\tiny\yng(1)}   \qquad
\longleftrightarrow \qquad
U(n_f-N_c)_{-k, \, -k + l (n_f-N_c)}~, \;    n_f\, \overline{\tiny\yng(1)}~,}
\ee
Upon choosing a positive FI parameter, the Higgs branch of the electric theory is given by the complex Grassmannian ${\rm Gr}(N_c, n_f)$. This duality has a natural interpretation in terms of the obvious geometric isomorphism of the dual Higgs branches:
\be
{\rm Gr}(N_c, n_f) \cong {\rm Gr}(n_f-N_c, n_f)~.
\ee
In this Higgs phase, certain choices of the levels $k$, $l$ have an interpretation in terms of the (generalised) quantum $K$-theory of ${\rm Gr}(N_c, n_f)$ -- see~{\it e.g.} \cite{Kapustin:2013hpk, Jockers:2019lwe, Ueda:2019qhg, Gu:2020zpg, Jockers:2021omw, Gu:2022yvj}. We will come back to this point in future work. 

\subsection*{Acknowledgements}

We thank Mathew Bullimore, Stefano Cremonesi, Heeyeon Kim, Horia Magureanu, Sakura Schafer-Nameki, and Eric Sharpe for discussions and correspondence. 
CC is a Royal Society University Research Fellow and a Birmingham Fellow, and his work is supported by the University Research Fellowship Renewal 2022 `Singularities, supersymmetry and SQFT invariants'. The work of OK is supported by the School of Mathematics at the University of Birmingham.

\appendix

\section{Real mass deformations and the $l=0$ dualities} \label{appendix: defs}
In this appendix, we briefly review the derivation of the minimally, marginally and maximally chiral dualities for $l=0$, assuming Aharony duality, as originally discussed in~\cite{Benini:2011mf}. We revisit this analysis here for completeness. This also allows us to explain how to carry out this standard computation in the gauge-symmetry-preserving conventions spelled out in section~\ref{subsec:bareCS}.

\subsection{Integrating out massive chiral multiplets}
Consider a 3d $\CN=2$ chiral multiplet $\Phi$ coupled to $U(1)_I$ vectors multiplets with charges $Q_I$. We are interested in giving a large real mass $m_0 \in \R$ to $\Phi$, thus integrating it out from the description. In the limit  $|m_0| \rightarrow \infty$, the UV contact terms are shifted according to:
\be
\delta \kappa_{IJ} = \half Q_I Q_J  \sign(m_0)~.
\ee
This is easily generalised to any non-abelian symmetry that $\Phi$ might be charged under. Integrating out a massive chiral multiplet $\Phi$ always shifts the CS contact terms according to:
\be
\delta \kappa = - \kappa^\Phi \sign(m_0)~,
\ee
for any symmetry, where the contributions $\kappa^\Phi$ from a single chiral multiplet are defined as in section~\ref{subsec:bareCS}. Importantly, this means that the bare CS levels $K$ are shifted as:
\be
\delta K = \begin{cases} 0 \quad & \text{if}\; m_0>0~, \\
2 \kappa^\Phi  \quad & \text{if}\; m_0<0~.\end{cases}
\ee
Take for example a massive chiral multiplet $\Phi$ of charge $Q_F\in \Z$ under some $U(1)_F$ symmetry, with a bare CS level $K_F$. After integrating out $\Phi$, we have the $U(1)_F$  bare CS level $K_F'= K_F$ if $m_0 \rightarrow \infty$ and $K_F'= K_F-Q_F^2$ if $m_0 \rightarrow - \infty$. Note that the bare Chern-Simons levels so obtained are integer-quantised, as needed by gauge invariance.

\subsection{Flowing from Aharony duality}
Let us consider the `electric' theory in Aharony duality, $U(N_c)_0$ with $N_f$ fundamentals and $N_f$ antifundamentals. We can obtain any SQCD$[N_c, k, 0, n_f, n_a]$ (with $l=0$) by appropriately decoupling flavours.

\medskip
\noindent
Let us then consider a particular RG flow:
\be
\delta \; : \; \text{SQCD$[N_c, 0, 0, N_f, N_f]$} \; + \; \delta m_0  \qquad \rightarrow \qquad \text{SQCD$[N_c, k, 0, n_f, n_a]$}~,
\ee
which is triggered by a particular choice of real mass $m_0$ in the UV theory. 
 This deformation will generate various flavour and mixed gauge-flavour CS levels $K_{GF}^{(e)}$ and $\Delta K_{FF}^{(e)}$, schematically. The $K_{GF}^{(e)}$ levels are part of the definition of the infrared theory. The pure flavour CS levels corresponds to local terms which can be removed at will. Recall that we choose $K_{FF}=0$ as part of our definition of unitary SQCD.

Because any real mass term is a VEV of a background vector multiplet, we can easily identify the dual mass deformation in the Aharony dual theory reviewed in section~\ref{subsec:Ah dual}. By following the corresponding RG flow, we arrive at a dual description for SQCD$[N_c, k, 0, n_f, n_a]$:
\be
\delta \; : \; \text{dual SQCD$[N_c, 0, 0, N_f, N_f]$}  \; + \; \delta m_0 \qquad \rightarrow \qquad \text{dual SQCD$[N_c, k, 0, n_f, n_a]$}~,
\ee
In this process, we again generate CS levels $K_{GF}^{(m)}$ and $\Delta K_{FF}^{(m)}$. The bare flavour CS levels shifts are to be added to the levels \eqref{aharony-bare-cs-levels-1}-\eqref{Aharony-cs-levels} encountered in the Aharony dual theory. We then compute the magnetic flavour CS levels of dual SQCD as:
\be
K_{FF}= K^{(\text{Ah})}+ \Delta K_{FF}^{(m)}- \Delta K_{FF}^{(e)}~,
\ee
where we shifted the `electric' flavour CS levels generated by the RG flow to the `magnetic' side of the duality.

\subsubsection{Minimally chiral duality: $|k|>|k_c|$}
\begin{table}[t]
\renewcommand{\arraystretch}{1.1}
\centering
\be\nn
\begin{array}{|c|c|cc|ccc|c|}
    \hline
        &  U(N_c)  &  SU(n_f)  &  SU(n_a)   &  U(1)_A  &  U(1)_T  &  U(1)_R & U(1)_0 \\
    \hline
    Q_i        &{\tiny\yng(1)} & \overline{\tiny\yng(1)} & \bm{1}& 1   & 0   &r &0 \\
\tilde{Q}^j   & \overline{\tiny\yng(1)} & \bm{1} &{\tiny\yng(1)}  & 1   & 0   &r &0\\
        \hline
     Q_\alpha        &{\tiny\yng(1)} & \overline{\tiny\yng(1)} & \bm{1}& 1   & 0   &r&\epsilon \\
\tilde{Q}^\beta   & \overline{\tiny\yng(1)} & \bm{1} &{\tiny\yng(1)}  & 1   & 0   &r&\epsilon \\
    \hline
    \end{array}
\ee
    \caption{Fields and charges for the mass deformation of the electric theory in the minimally chiral case. Every field charged under  $U(1)_0$ is integrated out. Here, we use the flavour indices $i= 1, \cdots, n_f$, $j=1, \cdots, n_a$, $\alpha=1,\cdots, p$, and $\beta=1,\cdots, q$, as well as $\epsilon = \sign(m_0)$. We only keep track of the flavour symmetries that survive in the infrared (and of $U(1)_0$).}\label{table: Ah elec flow 1}
\end{table}
To derive the minimally chiral Seiberg-like duality from Aharony duality, we start with the electric theory SQCD$[N_c, 0,0, N_f, N_f]$ and we integrate out $p$ fundamental and $q$ antifundamental chiral multiplets with a common real mass $m_0$, so that we obtain SQCD$[N_c, k,0, n_f, n_a]$ with:
\be
n_f = N_f-p~, \qquad n_a = N_f-q~, \qquad k= \half(p+q) \sign(m_0)~.
\ee
Note that $k_c\equiv \half (n_f-n_a)= \half(q-p)$ and $|k|= \half(p+q)$, hence we flow to minimally chiral SQCD, $|k| > |k_c|$, if and only if $p>0$ and $q>0$. The light and heavy fields are shown in table~\ref{table: Ah elec flow 1}. We identify some symmetry $U(1)_0$ along which we deform, with $\epsilon \equiv \sign(m_0)= \sign(k)$. 
\begin{table}[t]
\renewcommand{\arraystretch}{1.1}
\centering
\be\nn
\begin{array}{|c|c|cc|ccc|c|}
    \hline
        &  U(N_c^{\text{D}})  &  SU(n_f)  &  SU(n_a)   &  U(1)_A  &  U(1)_T  &  U(1)_R & U(1)_0 \\
    \hline
     q_j & \square  &\textbf{1}&  \overline{\square} &  -1 & 0 & 1-r & 0 \\
        \tilde{q}^i & \overline{\square}  &  \square  &  \textbf{1} &  -1 & 0 & 1-r & 0 \\
      {M_i}^j & \textbf{1} & \overline{\square} &  \square & 2 &0& 2r &0\\
     \hline
       q_\beta & \square  &  \textbf{1}  &  \textbf{1} &  -1 & 0 & 1-r & -\epsilon \\
        \tilde{q}^\alpha & \overline{\square} &  \textbf{1}  &  \textbf{1} &  -1 & 0 & 1-r & -\epsilon \\
      {M_i}^\beta & \textbf{1} & \overline{\square}&  \textbf{1} & 2 &0& 2r &\epsilon\\
      {M_\alpha}^j & \textbf{1} & {\textbf{1}} &  \square &  2 &0& 2r &\epsilon\\
      {M_\alpha}^\beta & \textbf{1} & {\textbf{1}} &  \textbf{1} &  2 &0& 2r &2\epsilon\\
      \mathfrak{T}^+  &  \textbf{1} & \textbf{1} & \textbf{1} & -N_f & 1 & r_T & -p \; (\epsilon=1) \; \text{or}\;  q\;  (\epsilon=-1)   \\
       \mathfrak{T}^-  &  \textbf{1} & \textbf{1} & \textbf{1} & -N_f & -1 & r_T & -q\;  (\epsilon=1)  \; \text{or}\;  p\;  (\epsilon=-1)   \\
      \hline
    \end{array}
    \ee
    \caption{Fields and charges for the mass deformation of the Aharony dual theory, in the minimally chiral case. Here $\epsilon= \sign(k)$.  }\label{table: Ah magn flow 1}
\end{table}
For $k_c\neq 0$, we also need to shift the origin of the real Coulomb branch and of the FI parameter according to:
\be
\sigma_a \rightarrow \sigma_a + {k_c\ov N_f} m_0~, \qquad \qquad
\xi \rightarrow \xi - k_c |m_0|~.
\ee
In other words, the $U(1)_0$ symmetry mixes with the gauge symmetry and with the topological symmetry~\cite{Benini:2011mf}.

\medskip
\noindent
On the magnetic side, we consider the Aharony dual gauge theory $U(N_c^D)_0$ with rank:
\bea
	&N_c^D &=&\; N_f - N_c \\ &&=&\; |k| + \frac{1}{2}(n_f + n_a)-N_c~. 
\eea
The $U(1)_0$ charge assignment in the dual theory are shown in table~\ref{table: Ah magn flow 1}. 
For $p$ and $q$ strictly positive, all the fields in the bottom rows must be integrated out. 
The bare CS levels are shifted according to:
\be
K_{IJ}\rightarrow K_{IJ} - \sum_{\Phi \, | \, Q_0[\Phi]<0} Q_I[\Phi]Q_J[\Phi]~,
\ee
where the sum is over all chiral multiplets with strictly negative $U(1)_0$ charge. 
 To determine the $U(1)_0$ assignement of the gauge singlets $\mathfrak{T}^{\pm}$, one recalls that they are identified with the monopoles of the electric theory. The latter have an induced $U(1)_0$ charge given by: 
\begin{equation}
	{Q}_0 \left[\mathfrak{T}^{\pm}\right]= \mp   \half \sum_{\psi} Q_G[\psi] \Big|Q_0[\psi]\Big|  - \frac{1}{2}\sum_{\psi} {Q}_0[\psi]~,
\end{equation}
where $Q_G$ denotes the gauge charge and we sum over all Dirac fermions in the theory.
It is then a straightforward exercise to derive the minimally chiral dual theory spelled out in section~\ref{subsec:l0minchir}.

\subsubsection{Marginally chiral duality: $|k|=|k_c|$}

The marginally chiral cases can be obtained as special limits of the minimally chiral case, when either $p$ or $q$ vanishes:
\bea
& k=k_c >0 \quad &&\Leftrightarrow \quad  p=0~, \quad \epsilon =1~,\\
& k=-k_c >0 \quad &&\Leftrightarrow \quad  q=0~, \quad \epsilon =1~,\\
& k=k_c <0 \quad &&\Leftrightarrow \quad  p=0~, \quad \epsilon =-1~,\\
& k=-k_c <0 \quad &&\Leftrightarrow \quad  q=0~, \quad \epsilon =-1~.
\eea 
In those cases, we see from~\ref{table: Ah magn flow 1} that either $\T^+$ or $\T^-$ survives the mass deformation of the dual theory.

\begin{table}[t]
\renewcommand{\arraystretch}{1.1}
\centering
\be\nn
\begin{array}{|c|c|cc|ccc|c|}
	\hline
	 & U(N_c)&SU(n_f)&SU(n_a)&U(1)_A&U(1)_T&U(1)_R&U(1)_0\\
	 \hline
	 Q_i &\square & \overline{\square}&\textbf{1}& 1& 0 & r & 0\\
	 \tilde{Q}^j & \overline{\square} & \textbf{1} & \overline{\square} & 1 & 0 & r & 0\\
	 \hline
	 \tilde{Q}^\gamma & \overline{\square} & \textbf{1} & \textbf{1} & 1 & 0 &r & 1\\
	 \tilde{Q}^\delta & \overline{\square} & \textbf{1} & \textbf{1} & 1 & 0 & r & -1\\
	 \hline
	 {Q}_\gamma &  {\square} & \textbf{1} & \textbf{1} & 1 & 0 &r & 1\\
	  {Q}_\delta & {\square} & \textbf{1} & \textbf{1} & 1 & 0 & r & -1\\
	 \hline
  \end{array}
    \ee
	\caption{Fields and charges for the mass deformation of the Aharony electric theory which leads to SQCD with $|k|<|k_c|$, with the massive fields for either $k_c>0$ (middle two rows) or $k_c<0$ (bottom two rows). Here $\gamma=1, \cdots, q$ and $\delta=1, \cdots, \t q$.}\label{elec charges mass def to max}
\end{table}

\subsubsection{Maximally chiral duality: $|k|<|k_c|$}
\begin{table}[t]
\renewcommand{\arraystretch}{1.1}
\centering
\be\nn
\begin{array}{|c|c|cc|ccc|c|}
	\hline
		 & U(N_c^\text{D})  & SU(n_f)& SU(n_a) & U(1)_A & U(1)_T&U(1)_R &U(1)_0  \\
		 \hline
	     q_j & \square & \textbf{1} & \overline{\square} & -1 & 0 & 1-r & 0\\
		   \tilde{q}^{i} & \overline{\square} & \square & \textbf{1}& -1 & 0 & 1-r  & 0\\
		{M_i}^j& \textbf{1} & \overline{\square} & \square& 2 & 0 & 2r &0\\
		 \hline
		q_\gamma & \square & \textbf{1}&\textbf{1}& -1 &0 & 1-r & -1 \\ 
		  q_{\delta} & \square & \textbf{1}&\textbf{1}& -1 & 0 & 1-r &1 \\ 
		 {M_i}^\gamma & \textbf{1} &\overline{\square} & \textbf{1} & 2 & 0 & 2r & 1\\
		 {M_i}^\delta & \textbf{1} & \overline{\square}& \textbf{1}   & 2 & 0 & 2r & -1\\
		 \mathfrak{T}_+ & \textbf{1}& \textbf{1}& \textbf{1}&-N_f & 1 & r_T &\t q\\ 
		 \mathfrak{T}_- & \textbf{1}& \textbf{1}& \textbf{1}&-N_f & -1 & r_T & -q\\ 
		 \hline
		  \tilde{q}^\gamma & \square & \textbf{1}&\textbf{1}& -1 &0 & 1-r & -1 \\ 
		 \tilde{q}^{\delta} & \square & \textbf{1}&\textbf{1}& -1 & 0 & 1-r &1 \\ 
		 {M_\gamma}^j & \textbf{1} & \textbf{1} & \square & 2 & 0 & 2r & 1\\
		 {M_\delta}^j & \textbf{1} & \textbf{1}&\square & 2 & 0 & 2r & -1\\
		 \mathfrak{T}_+ & \textbf{1}& \textbf{1}& \textbf{1}&-N_f & 1 & r_T & - q\\ 
		 \mathfrak{T}_- & \textbf{1}& \textbf{1}& \textbf{1}&-N_f & -1 & r_T & \t q \\ 
		 \hline
 \end{array}
    \ee
    	\caption{Fields and charges for the mass deformation of the Aharony dual theory, to obtain the maximally chiral case. The middle rows are for $k_c>0$, and the bottom rows for $k_c<0$.}\label{charges def to max chiral}
\end{table}
In the maximally chiral case, we start from the electric side of Aharony duality and integrate fundamental or antifundementals with opposite masses. Consider first the case $k_c >0$. In this case, we choose to integrate out $q$ antifundamental with a positive mass, and $\t q$ antifundamental with negative mass, so that we obtain:
\be
n_f= N_f~, \qquad n_a= N_f -q-\t q~, \qquad k= \half(q-\t q)~.
\ee
Note that $k_c=\half(q+\t q)$ in this case, hence $|k|< k_c$ as expected. We also need to shift the Coulomb branch origin and the the FI term according to:
\be
\sigma_a \rightarrow \sigma_a + {k\ov N_f} m_0~, \qquad \qquad
\xi \rightarrow \xi - k_c |m_0|~.
\ee
 For $k_c<0$,  we similarly choose to integrate out fundamental multiplets as shown in table~\ref{elec charges mass def to max}. Then we have: 
.\be
n_f= N_f -q-\t q~, \qquad n_a= N_f~, \qquad k= \half(q-\t q)~.
\ee
with $k_c=-\half(q+\t q)$. In either case, the dual gauge group is $U(N_c^D)_{-k}$ with $N_c^D= \max(n_f, n_a)-N_c$.  The details of the dual theory can be worked out from the charge assignment shown in table~\ref{charges def to max chiral} for the Aharony dual fields.

\bibliography{3dbib}

\bibliographystyle{JHEP}
\end{document}